\documentclass[a4paper,11pt]{article}
\usepackage{amssymb,graphicx,amsmath}

\makeatletter 
\renewcommand\section{\@startsection{section}{1}{\z@}              {-3.25ex\@plus -1ex \@minus -.2ex}                                    {1.5ex \@plus .2ex}                                    {\reset@font\Large\bfseries}}
\renewcommand\subsection{\@startsection{subsection}{2}{\z@}                                    {3.25ex \@plus 1ex \@minus.2ex}                                    {-1em}                                    {\reset@font\large\bfseries}}
\@addtoreset{equation}{section} \makeatother
\renewcommand{\theequation}{\thesection.\arabic{equation}}
\input{tcilatex}
\begin{document}

\setcounter{page}{0} \topmargin0pt \oddsidemargin0mm \renewcommand{%
\thefootnote}{\fnsymbol{footnote}} \newpage \setcounter{page}{0}
\begin{titlepage}
\indent \\
\vspace{1cm}
\begin{center}
{\Large {\bf Form factors of boundary fields for $A_{2}$-affine
Toda field theory}}

\vspace{1cm} {\large  \text{Olalla A.~Castro-Alvaredo}}

\vspace{0.4cm}
{ Centre for Mathematical Science, City University London, \\
Northampton Square, London EC1V 0HB, UK}
\end{center}
\vspace{1.5cm}

In this paper we carry out the boundary form factor program for
the $A_2$-affine Toda field theory at the self-dual point. The
latter is an integrable model consisting of a pair of particles
which are conjugated to each other, that is $1=\bar{2}$, and
possessing two bound states resulting from the scattering
processes $1 +1 \rightarrow 2$ and $2+2 \rightarrow 1$. We obtain
solutions up to four particle form factors for two families of
fields which can be identified with spinless and spin-1 fields of
the bulk theory. Previously known as well as new bulk form factor
solutions are obtained as a particular limit of ours. Minimal
solutions of the boundary form factor equations for all
$A_n$-affine Toda field theories are given, which will serve as
starting point for a generalisation of our results to higher rank
algebras.

\vspace{1cm}


\vfill{ \hspace*{-9mm}
\begin{tabular}{l}
\rule{6 cm}{0.05 mm}\\
$\text{o.castro-alvaredo@city.ac.uk}$\\
\end{tabular}}

\renewcommand{\thefootnote}{\arabic{footnote}}
\setcounter{footnote}{0}

\end{titlepage}
\newpage
\section{Introduction}
In the context of $1+1$-dimensional integrable quantum field
theories (IQFTs), form factors are defined as tensor valued
functions representing matrix elements of some local operator
$\mathcal{O}({x})$ located at the origin $x=0$ between a
multi-particle {\em{in}}-state and the vacuum:
\begin{equation}
F_{n}^{\mathcal{O}|\mu _{1}\ldots \mu _{n}}(\theta _{1},\ldots
,\theta _{n}):=\left\langle
0|\mathcal{O}(0)|\theta_1,\ldots,\theta_n\right\rangle_{\mu_1,\ldots,\mu_n}^{\text{in}}
~.\label{fofa}
\end{equation}
Here $|0\rangle$ represents the vacuum state and
$|\theta_1,\ldots,\theta_n\rangle_{\mu_1,\ldots,\mu_n}^{\text{in}}$
the physical ``in'' asymptotic states. The latter carry indices
$\mu_i$, which are quantum numbers characterizing the various
particle species, and depend on the real parameters $\theta_i$,
which are called rapidities. The form factors are defined for all
rapidities by analytically continuing from some ordering of the
latter; a fixed ordering provides a complete basis of states.

The form factor program for integrable models was pioneered by
P.~Weisz and M.~Karowski \cite{Weisz,KW} in the late 70s and
thereafter developed to a large extent by F.~A.~Smirnov, who also
formulated some of the consistency equations for form factors and
noticed that they corresponded to a particular deformation of the
Knizhnik-Zamolodchikov equations \cite{smirnovbook}. It was found
that the form factors of local operators can be obtained as the
solutions to a set of consistency equations which characterize the
analyticity and monodromy properties of the form factors and
constitute what is known as a Riemann-Hilbert problem. It is
therefore a priori possible, by solving these equations, to
compute all $n$-particle form factors associated to any local
field of a massive IQFT. This is however a very difficult
mathematical problem which has only been fully completed for free
theories. However, even partial solutions to the problem have
proven over the years to be extremely useful for the computation
of physical quantities. The reason is that all these quantities
are related in a way or another to correlation functions of local
fields. At the same time, it is a very well known fact that
correlation functions can be expressed as infinite sums depending
on the form factors of the fields involved. Although such sums can
not be performed analytically, they happen to converge very
quickly in terms of the number of particles of the form factors
involved. Indeed, convergence is often so good that the knowledge
of the two particle form factors is enough to get extremely
precise results (see e.g. \cite{Z,FMS}). This is one of the main
reasons why, since it first appeared, the bulk form factor program
has attracted so much attention and has been employed successfully
for the computation of correlation functions of many integrable
models (see \cite{Essler:2004ht} for a recent review).

A natural generalisation of these ideas is their extension  to
integrable theories with boundaries. The study of IQFTs with
boundaries goes back to the works of I.~V.~Cherednik
\cite{Cherednik:1985vs} and E.~V.~Sklyanin \cite{Sklyanin:1988yz},
where the conditions for a boundary to preserve integrability were
established in the form of generalized Yang-Baxter equations
which, given the bulk $S$-matrix, can be solved for the reflection
amplitudes off the boundary. These pioneering works were followed
by \cite{Fring:1993wt,Fring:1994ci,Fring}, where the authors were
concerned with the computation of reflection amplitudes in ATFTs,
and by \cite{Ghoshal:1993tm}, where the crossing property of the
reflection amplitudes as well as an explicit realization of the
boundary were proposed. The latter work provided in fact the first
tool for the computation of form factors in the presence of a
boundary. The main idea is  to realize the boundary as a
``boundary state" located at the origin of time ($t=0$) and
expressible in terms of the same creation-annihilation operators
used to build the bulk Hilbert-space. The drawback of this
approach is that the boundary state is in fact an infinite sum of
particle states of the bulk theory, which makes computations very
involved, even for free theories \cite{Konik:1995ws,dirk}. It is
worth saying that within this approach, only form factors of
fields sufficiently far from the boundary can be obtained, e.g.
form factors of fields sitting at the boundary are not accessible.
A complementary approach which allows the computation of form
factors of fields sitting at the boundary has been proposed
recently \cite{BPT}. It turns out that if one considers the
boundary as a point-like object sitting at the origin of space
($x=0$), then a natural generalisation of the bulk form factor
program for IQFTs is possible. We will call this the boundary form
factor program. Within this program the form factors of fields
sitting at the boundary can be computed for particular models by
solving a Riemann-Hilbert problem which is reminiscent of the bulk
case. Indeed, as for the bulk case, this Riemann-Hilbert problem
is very similar to the set of equations employed previously for
the computation of the form factors of integrable spin chains with
boundaries \cite{Jimbo:1994gm}-\cite{Furutsu:1999by}. The latter
equations result from a generalisation of the approach developed
by M.~Jimbo, T.~Miwa and their collaborators \cite{JMi,JMi2,JMi3}
for bulk models. The first form factors obtained by means of the
boundary form factor program were found in \cite{BPT} for several
theories with a single-particle spectrum. An extension of this
work followed in \cite{mypaper} where higher particle form factors
of the sinh-Gordon model were obtained and in \cite{st} where a
precise counting of the number of solutions to the boundary form
factor equations of the sinh-Gordon and Yang-Lee models was
performed. In this paper we intend to extend this program to a
class of multi-particle theories: the $A_n$-affine Toda field
theories.

Affine Toda field theories (ATFTs) have been studied since a long
time and have played a prominent role in the development of the
field of integrable field theories \cite{toda1,toda2}. Notably,
the picture proposed by A.~B. Zamolodchikov whereby IQFTs may be
regarded as conformal field theories perturbed by relevant fields
was first illustrated on the example of the $E_8$-ATFT
\cite{P2CFT, PCFT}. From this example and from extended work on
classical Toda theory it was expected that a different theory
should exist for each simple Lie algebra and that this Lie
algebraic structure would be crucial in the understanding of these
models. Subsequently, a lot a work was carried out in order to
compute the $S$-matrices of ATFTs related to each simple Lie
algebra. Much of this work was reviewed and extended in
\cite{BCDS}, where expressions for the $S$-matrices of all ATFTs
can be found. More recently a closed universal formula for the
$S$-matrices of all ATFTs which depends solely on Lie algebraic
quantities has been obtained \cite{FKS}.

Amongst the large class of ATFTs the $A_n$-case is probably the
one for which structures are most simple. For example,
two-particle scattering amplitudes have poles in the physical
sheet whose order is at most two (in contrast to other ATFTs where
poles of up to order 12 occur). As a consequence the pole
structure of both bulk \cite{oota,Babujian:2003za} and boundary
form factors is less involved. The $S$-matrices of the
$A_n$-series were first obtained in \cite{asmatrices}, where it
was suggested that they should be given by the $S$-matrices of the
corresponding minimal $A_n$-Toda field theories (higher rank
generalisations of the Ising model, then known as
$\mathbb{Z}_n$-models) obtained originally in
\cite{Koberle:1979sg}, multiplied by a CDD factor which contained
the dependence on the coupling constant.

In the specific context of ATFTs with boundaries, a great deal of
work has been carried out in the last years to compute the
corresponding reflection probabilities and to classify the
boundary conditions that are consistent with integrability. The
first reflection probabilities for ATFTs associated to
simply-laced Lie algebras were obtained by A.~Fring and
R.~K{\"o}berle in a series of papers
\cite{Fring:1993wt,Fring:1994ci,Fring}. Notably, the last of these
papers dealt for the first time with the case of dynamic
boundaries. Another series of papers by members of the Durham and
York mathematics groups \cite{corrigan,corrigan3,corrigan2}
achieved the full classification of boundary conditions which are
compatible with integrability. Later on, many authors have
computed further reflection amplitudes for various ATFTs (a quite
complete list of references can be found in \cite{reflections}).
Given the topic of this paper, it is worth emphasizing those
contributions which dealt with the $A_n$-case. Besides the work
cited above, solutions for all $A_n$-ATFTs  in terms of hyperbolic
functions were obtained in \cite{corrigan,garden2}. The specific
$A_2$-case was studied in detail in \cite{garden1}. Integral
representations of the solutions obtained in \cite{corrigan}, as
well as solutions for other algebras were given in \cite{fateev}.
Finally, it is worth emphasizing that the $A_1$-case (that is, the
sinh-Gordon model) has, independently from other $A_n$-ATFTs,
attracted a lot of interest as its reflection amplitudes exhibit
some properties that single them out from those of all other
$A_n$-ATFTs \cite{sinh1}-\cite{sinh7}.

One of the main differences between the solutions in
\cite{Fring:1993wt,Fring:1994ci,Fring} and those in
\cite{garden1,garden2,fateev,corrigan} is the fact that whereas
the former are invariant under a duality transformation (we will
see later what this means) the latter are not. However it was
found in \cite{reflections} that these two types of solutions can
be related to each other in a simple fashion (see section
\ref{ref}). In fact that work provided the first closed solution
for the boundary reflection amplitudes which holds for all ATFTs
and which can be related via CDD factors to all known solutions in
the literature so far. In this paper we will take the reflection
amplitudes computed in \cite{corrigan, fateev} as  starting point.

The paper is organized as follows: In section \ref{s1} we review
the expressions and several properties of the $S$-matrices and
reflection amplitudes of the $A_n$-ATFTs. In section \ref{s2} we
introduce the boundary form factor program \cite{BPT} and compute
the minimal one- and two-particle form factors for all
$A_n$-ATFTs. We also give a general ansatz for the higher particle
form factors, thereby characterizing their pole structure. In
section \ref{s3} we concentrate on the derivation of recursive
equations for the form factors of $A_2$-ATFT and find solutions
for them at the self-dual point, $B=1$,  and up to four particles.
We show that some of our solutions correspond to the same class of
fields whose bulk form factors were computed by T.~Oota in
\cite{oota}. In addition, we also construct the form factors up to
four particles for another class of fields which correspond to
spin-1 operators of the bulk theory. In section \ref{s4} we
present our conclusions and outlook.

\section{The $A_{n}$-ATFTs : Generalities} \label{s1}

The $A_{n}$-ATFTs are a family of IQFTs whose exact $S$-matrices
\cite{asmatrices}, $R$-matrices
\cite{Fring:1993wt,corrigan,Fring:1994ci,Fring,garden1,garden2,fateev}
and mass spectrum \cite{Koberle:1979sg} are completely
characterized by an underlying Lie algebraic structure related to
the simply-laced algebra $A_{n}$. The theory has $n$ stable
particles in the spectrum with masses
\begin{equation}
    m_a=2m \sin\frac{a \pi}{h} \qquad a=1,\ldots,n,
\end{equation}
in terms of a fundamental mass scale $m$ and the Coxeter number of
$A_n$, $h=n+1$. If we arrange these masses into a vector
$(m_1,\ldots,m_n)$ then this is the Perron-Frobenius eigenvector
of the Cartan matrix of $A_{n}$. It is common to relate these
particles to the nodes of the Dynkin diagram of $A_{n}$. In
accordance with the symmetry of the latter, the anti-particle of a
particle $i$ corresponds to the $h-i$
node, as shown in the figure:\\

\unitlength=0.680000pt
\begin{picture}(437.92,75.00)(50.00,95.00)
\qbezier(260.00,140.00)(365.00,95.00)(471.00,140.00)
\qbezier(301.00,140.00)(365.00,110.00)(431.00,140.00)
\qbezier(341.00,140.00)(365.00,125.00)(391.00,140.00)
\put(396.00,165.00){\makebox(0.00,0.00){$\bar{3}$}}
\put(436.00,165.00){\makebox(0.00,0.00){$\bar{2}$}}
\put(476.00,165.00){\makebox(0.00,0.00){${n=\bar{1} }$}}
\put(336.00,165.00){\makebox(0.00,0.00){$3$}}
\put(296.00,165.00){\makebox(0.00,0.00){$2$}}
\put(255.00,165.00){\makebox(0.00,0.00){$1$}}
\put(400.00,150.00){\line(1,0){30.00}}
\put(440.00,150.00){\line(1,0){30.00}}
\put(390.00,150.00){\line(-1,0){10.00}}
\put(340.00,150.00){\line(1,0){10.00}}
\put(300.00,150.00){\line(1,0){30.00}}
\put(260.00,150.00){\line(1,0){30.00}}
\put(395.00,150.00){\circle*{10.00}}
\put(435.00,150.00){\circle*{10.00}}
\put(475.00,150.00){\circle*{10.00}}
\put(255.00,150.00){\circle*{10.00}}
\put(295.00,150.00){\circle*{10.00}}
\put(335.00,150.00){\circle*{10.00}}
\end{picture}
\begin{center}
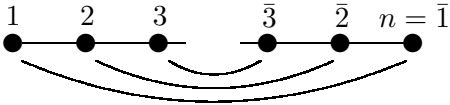
\emph{Figure 1: The Dynkin diagram of the simply-laced algebra
$A_n$}
\end{center}
\subsection{The $S$-matrices}\indent \\

\noindent The $S$-matrices admit both an integral
\cite{Oota:1997un,frenkel} and a block representation
 \cite{asmatrices}. The integral representation takes the form
\begin{equation}\label{s}
  S_{ab}(\theta,B)=\exp\left[ \int_{0}^{\infty} \frac{dt}{t }\, \Phi_{ab}(t,B)  \sinh\left(\frac{t \theta}{i
  \pi}\right)\right],
\end{equation}
with
\begin{equation}\label{phi}
\Phi_{ab}(t,B)=\frac{8\sinh\left(\frac{tB}{2 h}\right)
  \sinh\left(\frac{t(2-B)}{2 h}\right)\sinh\left(\frac{t\min(a,b)}{h}\right)
  \sinh\left(\frac{t(h-\max(a,b))}{h}\right)}{\sinh\left(\frac{t}{h}\right) \sinh t},
\end{equation}
where $B$ is  the effective coupling constant which is a function
of the coupling constant $\beta$ which appears in the classical
Lagrangian of the theory . By construction the $S$-matrices are
invariant under the transformation $B \rightarrow 2-B$, which is
equivalent to week-strong duality, $\beta \rightarrow 4 \pi/\beta$
of the Lagrangian \cite{BCDS}. The point $B=1$ is known as the
self-dual point. The kernel above admits an alternative
representation, based on the identity
\begin{equation}
\frac{\sinh\left(\frac{t\min(a,b)}{h}\right)
  \sinh\left(\frac{t(h-\max(a,b))}{h}\right)}{\sinh\left(\frac{t}{h}\right) }=\sum_{\begin{tabular}{c}
   \scriptsize{ $p=|a-b|+1$} \\
    \text{\scriptsize step 2} \\
  \end{tabular}}^{a+b-1} \sinh\left(\frac{t(h-p)}{h}\right),
\end{equation}
which allows us to bring the $S$-matrix into the block product
form
\begin{equation}
    S_{ab}(\theta,B)=\prod_{\begin{tabular}{c}
   \scriptsize{ $p=|a-b|+1$} \\
    \text{\scriptsize step 2} \\
  \end{tabular}}^{a+b-1} \{p\}_{\theta},
\end{equation}
with blocks
\begin{equation}
    \{x\}_{\theta}=\frac{(x+1)_{\theta}(x-1)_{\theta}}{(x+1-B)_{\theta}
    (x-1+B)_{\theta}}, \qquad \text{and} \qquad (x)_\theta = \frac{\sinh\frac{1}{2}\left(\theta+ \frac{i \pi x}{h}\right)}
    {\sinh\frac{1}{2}\left(\theta- \frac{i \pi x}{h}\right)},
\end{equation}
or
\begin{equation}
    \{x\}_{\theta}=\exp\left[8 \int_{0}^{\infty} \frac{dt \sinh\left(\frac{tB}{2 h}\right)
  \sinh\left(\frac{t(2-B)}{2 h}\right) \sinh t \left(1- \frac{x}{h}\right)}{t \sinh t }  \sinh\left(\frac{t \theta}{i
  \pi}\right)\right].
\end{equation}
From the block form above it is easy to see that these
$S$-matrices possess simple and double poles. As usual in this
context, simple poles are related to the presence of bound states
in the theory which result from fusing processes of the form $a+b
\rightarrow \bar{c}$ whenever $a+b+c$  equals either $h$ or $2h$.
If such fusing occurs the amplitude $S_{ab}(\theta)$ will have a
pole $\theta=i u_{ab}^{\bar{c}}$ located at
\begin{equation}\label{thetaabc}
u_{ab}^{\bar{c}}=\left\{%
\begin{array}{ll}
    \frac{\pi (a+b)}{h} & \text{if} \quad a+b+c=h \\
   \frac{\pi (2h-a-b)}{h} & \text{if} \quad a+b+c=2h \\
\end{array}%
\right.\,.
\end{equation}
Closed formulae for the associated $S$-matrix residua, which are
closely related to the  on-shell three-point coupling
$\Gamma_{ab}^{\bar{c}}$, can be easily obtained. For example, for
the process $1+1 \rightarrow 2$:
\begin{equation}
   (\Gamma_{11}^{2})^2= -i \lim_{\theta \rightarrow i u_{11}^{2}}(\theta-i u_{11}^{2})
    S_{11}(\theta)=\frac{2 \sin\left(\frac{2\pi}{h}\right)\sin\left(\frac{(2-B)\pi}{2h}\right)\sin\left(\frac{B\pi}{2h}\right)}
    {\sin\left(\frac{(2+B)\pi}{2h}\right)\sin\left(\frac{(4-B)\pi}{2h}\right)}.
\end{equation}
\subsection{The reflection
amplitudes}\label{ref}\indent \\

\noindent A particular set of solutions for the $R$-matrices of
$A_{n}$-ATFTs was constructed in \cite{corrigan} in a block
product representation,
\begin{equation}
R_a(\theta,B)=R_{\bar{a}}(\theta,B)=\prod_{p=1}^{a}\|p\|_{\theta},
\end{equation}
with
\begin{equation}\label{b1}
\|x\|_{\theta}=\frac{\left(\frac{x-1}{2}\right)_{\theta}\left(\frac{x+1}{2}-h\right)_{\theta}}
{\left(\frac{x-1+B}{2}-h\right)_{\theta}\left(\frac{x+1-B}{2}\right)_{\theta}}.
\end{equation}
The corresponding integral representation was given in
\cite{fateev}
\begin{equation}
    {R}_a(\theta,B)=R_{\bar{a}}(\theta,B)=\exp\left[\int_{0}^{\infty}\frac{dt}{t} {\rho}_a(t,B) \sinh \left(\frac{t \theta}{i
    \pi}\right)\right], \label{rho}
\end{equation}
with
\begin{eqnarray}
{\rho}_a(t,B)= {\rho}_{\bar{a}}(t,B)&=&\frac{8
\sinh\left(\frac{t(2-B)}{4h}\right)\sinh\left(\frac{t
(2h+B)}{4h}\right) \sinh\left(\frac{t a}{2h}\right)\sinh\left(\frac{t(h-a)}{2h}\right)}{\sinh\left(\frac{t}{2h}\right) \sinh {t}}\nonumber \\
  &=& \frac{8 \sinh\left(\frac{t(2-B)}{4h}\right)\sinh\left(\frac{t (2h+B)}{4h}\right)}{ \sinh {t}} \sum_{p=1}^{a}
\sinh\left(\frac{t(h-2p+1)}{2h}\right).\label{bo}
\end{eqnarray}
A feature of these solutions is that, contrary to the $S$-matrices
(\ref{s}), they are not invariant under the duality transformation
$B\rightarrow 2-B$. As a result, they have the property of having
different classical limits in the weak ($B \rightarrow 0$) and
strong ($B \rightarrow 2$) coupling limits, namely
\begin{equation}\label{limits}
    R_a(\theta,0)=\prod_{b=1}^{n} S_{ab}(\theta,1),\qquad
    R_a(\theta,2)=1.
\end{equation}
This means that in the classical limit the amplitudes
$R_i(\theta,0)$ correspond to a theory with fixed Dirichlet
boundary conditions whereas $R_i(\theta,2)$ correspond to a theory
with von Neumann boundary conditions \cite{corrigan,
corrigan3,corrigan2}. How the breaking of duality invariance
occurs can be explicitly seen by writing the corresponding
$R$-matrices as
\begin{equation}\label{rr}
   {R}_a(\theta,B)=\tilde{{R}}_a(\theta,B)\prod_{b=1}^{n}
    S_{ba}\left(\theta,1-\frac{B}{2}\right),
\end{equation}
where $\tilde{R}$ are the duality invariant $R$-matrices
constructed in \cite{reflections} and the identity (\ref{rr}) was
also found there.\footnote{The $R$-matrices constructed in
\cite{reflections} admit an integral representation of the form
(\ref{rho}) with $$\rho_j(t,B)= \frac{8
\sinh\left(\frac{t(2-B)}{4h}\right)\sinh\left(\frac{t
B}{4h}\right)\sinh\left(\frac{t(1-h)}{2h}\right)
   \sinh\left(\frac{t j}{2h}\right)\sinh\left(\frac{t(h-j)}{2h}\right)}{\sinh^2\left(\frac{t}{2h}\right)\sinh
   t}$$ Notice that this kernel differs from the one given in \cite{reflections} which contained several typos. }
\section{The boundary form factor consistency equations}\label{s2}
A set of consistency equations whose solutions are the form
factors of boundary fields of an IQFT was recently proposed in
\cite{BPT}. There the equations were presented for the special
case of a theory with a single particle species, but it is trivial
to write them for the general case. The first three relations
\begin{eqnarray}
  F_{n}^{\mathcal{O}| \ldots \mu_i \mu_{i+1}\ldots}(\ldots, \theta_i,
  \theta_{i+1},\ldots) &=& S_{\mu_i \mu_{i+1}}(\theta_{i}-\theta_{i+1}) F_{n}^{\mathcal{O}|\ldots \mu_{i+1} \mu_{i}\ldots}(\ldots,\theta_{i+1}, \theta_i,
  \ldots),\label{eq1}
    \\
 F_{n}^{\mathcal{O}| \mu_1 \ldots \mu_{n-1} \mu_n}(\theta_1, \ldots,\theta_{n-1}, \theta_n)&=&
 R_{\mu_n}(\theta_n) F_{n}^{\mathcal{O}| \mu_1 \ldots \mu_{n-1} \mu_n}(\theta_1, \ldots,
\theta_{n-1},-\theta_n),\label{eq2}\\
F_{n}^{\mathcal{O}| \mu_1 \mu_2 \ldots  \mu_n}(\theta_1,\theta_2
\ldots, \theta_n)&=&
 R_{\mu_1}(i\pi-\theta_1) F_{n}^{\mathcal{O}| \mu_1 \mu_2 \ldots \mu_n}(2 \pi i-\theta_1,\theta_2, \ldots,\theta_n),\label{eq3}
\end{eqnarray}
express the fact that exchanging two particles in a form factor
amounts to the scattering of those particles, which is
appropriately taken into account by the two-particle $S$-matrix.
In addition if one of the particles a form factor depends upon is
scattered off the boundary with probability $R_{\mu_i}(\theta_i)$,
the value of the form factor should remain unchanged, which is
expressed by the second equation. Finally, the particle might be
scattered off the boundary after a $2 \pi i$ rotation, which is
expressed by the last equation. For $n=1$ and $n=2$ the minimal
solutions to these equations (that is, solutions with no poles in
the physical sheet) will play a crucial role in the formulation of
a general ansatz for the $n$-particles form factors and details on
their computation will be provided in the coming sections.

In addition to these three equations we have two more relations
which fix the pole structure of the form factors: the kinematic
residue equation
\begin{eqnarray}
\lim_{\bar{\theta}_{0} \rightarrow
\theta_0}F_{n+2}^{\mathcal{O}|\bar{\mu} \mu \mu_1 \ldots \mu_n}
 (\bar{\theta}_0 + i \pi, \theta_0, \theta_1,\ldots,\theta_n)&=& i\left(1-\prod_{k=1}^{n}S_{\mu \mu_{k}}({\theta}_{0}-\theta_k)
 S_{\mu \mu_{k}}({\theta}_{0}+\theta_k)\right)\nonumber \\
 &\times&
  F_{n}^{\mathcal{O}| \mu_1 \ldots \mu_n}(\theta_1, \ldots, \theta_n),\label{eq4}
\end{eqnarray}
and the bound state residue equation
\begin{eqnarray}
 \lim_{\epsilon \rightarrow 0}\epsilon F_{n+2}^{\mathcal{O}|a b  \mu_1 \ldots \mu_n}
 ({\theta} + \frac{i u_{ab}^{c}}{2}+\epsilon, {\theta} - \frac{i u_{ab}^{c}}{2}-\epsilon,
 \theta_1,\ldots,\theta_n)= i\Gamma_{ab}^{c}
  F_{n+1}^{\mathcal{O}|c \mu_1 \ldots \mu_n}(\theta,\theta_1, \ldots, \theta_n),\label{eq5}
\end{eqnarray}
where particles $a$ and $b$ fuse to produce particle $c$ which
implies that the scattering amplitude $S_{ab}(\theta)$ has a pole
at $\theta=i u_{ab}^{c}$ with residue given by
$(\Gamma_{ab}^{c})^2$.

\subsection{The minimal one-particle form factors}\indent \\

\noindent As explained in \cite{BPT}, starting with the boundary
reflection amplitudes, it is possible to find a minimal solution,
$w_a(\theta)$, to the one-particle form factor
equations\footnote{From here onwards we will omit the explicit
dependence of the reflection amplitudes, $S$-matrix and form
factors in $B$.}
\begin{equation}\label{1pff}
   w_{a}(\theta)=R_{a}(\theta)   w_{a}(-\theta)\qquad\text{and}\qquad   w_{a}(\theta+ i \pi)=
   R_{a}(-\theta)  w_{a}(i \pi-\theta),
\end{equation}
which takes the form
\begin{equation}
  w_{a}(\theta)=g_{a}(\theta)g_{a}(i \pi-\theta),
\end{equation}
where $g_a(\theta)$ satisfies the equations
\begin{equation}
    g_a(\theta)=R_{a}(\theta) g_a(-\theta)\qquad\text{and}\qquad g_a(\theta)=
   g_a(2 \pi i-\theta).
\end{equation}
These are the same equations as for the two-particle minimal form
factor, with the $S$-matrix replaced by the reflection amplitudes.
Therefore they can be solved in a systematic way, namely
\begin{equation}
    g_a(\theta)=g_{\bar{a}}(\theta,B)=\exp\left[-\frac{1}{2}\int_{0}^{\infty}\frac{dt}{t \sinh t} \rho_a(t,B) \cosh t \left(1+\frac{i
    \theta}{\pi}\right)\right].\label{gj}
\end{equation}
We can write (\ref{gj}) as an infinite product of Euler's gamma
functions by employing the identities:
\begin{equation}
    \int_{0}^{\infty} \frac{dt}{t}\frac{\sinh(\alpha t)\sinh(\beta t) e^{-\gamma
    t}}{\sinh(u t)}=\frac{1}{2}\log\left[\frac{\Gamma\left(\frac{\alpha+\beta+\gamma+u}{2u} \right)
    \Gamma\left(\frac{-\alpha-\beta+\gamma+u}{2u} \right)}{\Gamma\left(\frac{-\alpha+\beta+\gamma+u}{2u} \right)
    \Gamma\left(\frac{\alpha-\beta+\gamma+u}{2u} \right)}
    \right],\label{id}
\end{equation}
and
\begin{equation}\label{sinh}
    \frac{1}{\sinh t }= 2 e^{-t} \sum_{k=0}^{\infty} e^{-2 k t}.
\end{equation}
We obtain
\begin{eqnarray}
g_a(\theta,B)=g_{\bar{a}}(\theta,B)=\prod_{p=1}^{a}\frac{\sinh\frac{1}{2}\left(\theta-\frac{i
\pi (2p-B)}{2h}\right)\cosh\frac{1}{2}\left(\theta-\frac{i \pi
(B+2p-2)}{2h}\right)\varrho^{1}_{p-1}(\theta)\varrho_{p}^{\frac{1}{2}}(\theta)}
{\sinh\frac{1}{2}\left(\theta-\frac{
 i\pi (p-1)}{h}\right)\cosh\frac{1}{2}\left(\theta-\frac{i\pi
p}{h}\right)\varrho_{\frac{2p-B}{2}}^{1}(\theta)
\varrho_{\frac{B+2p-2}{2}}^{\frac{1}{2}}(\theta)} ,
\end{eqnarray}
where
\begin{equation}
   \varrho_{x}^{n}(\theta)= \prod_{k=1}^{\infty}\frac{\Gamma\left(k+n-\frac{x}{2h}+\frac{i \theta}{2\pi} \right)
   \Gamma\left(k+n-\frac{x}{2h}-\frac{i \theta}{2\pi} \right)}
    {\Gamma\left(k-n+\frac{x}{2h}-\frac{i \theta}{2\pi}
    \right)\Gamma\left(k-n+\frac{x}{2h}+\frac{i \theta}{2\pi}
    \right)},
    \qquad x \in [0,2h),\quad n \in \mathbb{Q}.
\end{equation}
The following identity will be useful for later purposes:
\begin{eqnarray}\label{pr}
  w_{\bar{a}}(\theta+ i \pi)
     w_{a}(\theta)=\prod_{p=1}^{a} \frac{\varrho^{0}_{2p}(2\theta)\varrho^{1}_{2p-2}(2\theta)}
    {\varrho^{0}_{B+2p-2}(2\theta)\varrho^{1}_{2p-B}(2\theta)}.
\end{eqnarray}
\subsection{The two-particle form factors}\indent \\

\noindent The two-particle form factors are the solutions of the
two-particle form factor equations
\begin{equation}\label{mini}
    F_{2}^{ab}(\theta_{1},\theta_2)=S_{ab}(\theta)
    F_{2}^{ba}(\theta_2,\theta_1)= R_{b}(\theta_2)
    F_{2}^{ab}(\theta_1,-\theta_2)=R_{a}(i\pi-\theta_1)
    F_{2}^{ab}(2\pi i -{\theta}_{1},\theta_2).
\end{equation}
It was shown in \cite{BPT} that solutions to these equations take
the general form
\begin{equation}
    F_{2}^{ab}(\theta_1,\theta_2)=f_{ab}(\theta_{12})f_{ab}(\hat{\theta}_{12})
    w_a(\theta_{1})w_{b}(\theta_{2})\Phi(y_{1},y_{2}),
    \quad\text{with}\quad y_i=e^{\theta_i}+e^{-\theta_i}.
\end{equation}
Here $\theta_{12}=\theta_1-\theta_2$,
$\hat{\theta}_{12}=\theta_1+\theta_2$ and $f_{ab}(\theta)$ is the
bulk two-particle minimal form factor, that is a solution of the
equations
\begin{equation}
   f_{ab}(\theta)=S_{ab}(\theta)
    f_{ba}(-\theta)=f_{ab}(2\pi i-\theta),\label{mi}
\end{equation}
with no poles in the physical strip $\text{Im}(\theta)\in
[0,\pi]$. The function $\Phi(y_1,y_2)$ characterizes the
particular operator and is a symmetric function of the variables
$y_1, y_2$, which is the constraint given in \cite{BPT} for models
with a single particle spectrum. In particular, the minimal
boundary two-particle form factors correspond to
$\Phi(y_1,y_2)=1$. Solutions to (\ref{mi}) for $A_{n}$-ATFTs were
first found by T.~Oota \cite{oota} in the form of an infinite
product of $\Gamma$-functions. Such representation can be obtained
in the usual way: for an $S$-matrix which admits an integral
representation of the type (\ref{s}) it is easy to show that a
solution to (\ref{mi}) is given by
\begin{equation}\label{fmin}
   f_{ab}(\theta)=\exp\left[ -\frac{1}{2}\int_{0}^{\infty}
\frac{dt}{t \sinh t}\, \Phi_{ab}(t,B)  \cosh t
  \left(1+\frac{i \theta}{\pi}\right)\right].
\end{equation}
As for the one-particle form factors, we can use (\ref{id}) and
(\ref{sinh}) to express the form factor above as an infinite
product of gamma functions
\begin{eqnarray}
  && f_{ab}(\theta) = \prod_{\begin{tabular}{c}
   \scriptsize{ $p=|a-b|+1$} \\
    \text{\scriptsize step 2} \\
  \end{tabular}}^{a+b-1} \frac{\varrho_{p+1}^{1}(\theta)\varrho_{p-1}^{1}(\theta)}
    {\left\langle -p \right\rangle_{\theta}\varrho_{p+1-B}^{1}(\theta)\varrho_{p-1+B}^{1}(\theta)},\label{ff}
  \end{eqnarray}
where
\begin{equation}\label{p}
    \left\langle p \right\rangle_{\theta}=\frac{\sinh\frac{1}{2}\left(\theta+ \frac{i \pi
(p-1)}{h} \right)\sinh\frac{1}{2}\left(\theta+ \frac{i \pi
(p+1)}{h} \right)}{\sinh\frac{1}{2}\left(\theta+ \frac{i \pi
(p-1+B)}{h} \right)\sinh\frac{1}{2}\left(\theta+ \frac{i \pi
(p+1-B)}{h} \right)}.
\end{equation}
It is easy to check that the function inside the product above
equals the function $F_{p}^{\text{min}}(\theta)$ introduced by
T.~Oota in \cite{oota}, as it should be. A useful identity
involving the minimal form factors above is
\begin{equation}
f_{{a} b}(\theta)f_{\bar{a}b}(\theta + i
\pi)=\prod_{\begin{tabular}{c}
   \scriptsize{ $p=|a-b|+1$} \\
    \text{\scriptsize step 2} \\
  \end{tabular}}^{a+b-1} \left\langle p \right\rangle_{\theta},
\end{equation}
which was also obtained in \cite{oota}. In addition, it is easy to
show from (\ref{ff}) that
\begin{eqnarray}
  f_{\bar{a}a}(2\theta+ i\pi) =
\prod_{p=1}^{\min(a,\bar{a})}\frac
   {\varrho^{0}_{B+2p-2}(2\theta)\varrho^{0}_{2p-B}(2\theta)}
    {\varrho^{0}_{2p}(2\theta)\varrho^{0}_{2p-2}(2\theta)}.
 \end{eqnarray}
Combining this with (\ref{pr}) we find that
\begin{equation}
    f_{\bar{a}a}(2\theta+ i\pi)
    w_{\bar{a}}(\theta+ i
    \pi)w_{{a}}(\theta)=\prod_{p=1}^{\min(\bar{a},a)}\frac{\sinh\left(\theta+\frac{i \pi(p-1)}{h} \right)
    \sinh\left(\theta-\frac{i \pi(p-1)}{h} \right)}{\sinh\left(\theta+\frac{i \pi(B-2p)}{2h} \right)\sinh\left(\theta-\frac{i \pi(B-2p)}{2h}
    \right)}.
\end{equation}
Also
\begin{equation}
    \frac{f_{11}(\theta +\frac{i \pi}{h} )f_{11}(\theta- \frac{i \pi}{h})}{f_{21}(\theta)}=
    \frac{\sinh\frac{1}{2}\left(\theta + \frac{i \pi}{h} \right)\sinh\frac{1}{2}\left(\theta -\frac{i \pi}{h} \right)}
    {\sinh\frac{1}{2}\left(\theta + \frac{i \pi(1- B)}{h} \right)\sinh\frac{1}{2}\left(\theta - \frac{i \pi (1-B)}{h}
    \right)}.\label{also}
\end{equation}
These identities will be very important in order to bring the
kinematic and bound state residue equations into a simple form.
\subsection{Higher particle form
factors}\indent \\

\noindent For reflection matrices which have no poles at $\theta=i
\pi/2$ (as is our case here) it is natural to make the following
ansatz for the $n$-particle form factors
\begin{eqnarray}
    F_{n}^{\mathcal{O}| \mu_1 \ldots \mu_n}(\theta_1, \ldots,
    \theta_n)&=& H_{n}^{\mathcal{O}|\mu_1 \ldots \mu_n} Q_{n}^{\mathcal{O}| \mu_1 \ldots \mu_n}(y_1,
    \ldots,y_n)\prod_{k=1}^{n}w_{\mu_k}(\theta_k) \nonumber \\
    &&\times \prod_{1\leq i<j \leq
    n}\frac{f_{\mu_i\mu_j}(\theta_{ij})f_{\mu_i\mu_j}(\hat{\theta}_{ij})}
    {P_{\mu_i\mu_j}(\theta_{ij})P_{\mu_i\mu_j}(\hat{\theta}_{ij})
    (y_i+y_j)^{\delta_{\mu_i \bar{\mu}_j}}}, \label{ansatz}
\end{eqnarray}
where $H_n^{\mathcal{O}|\mu_1 \ldots \mu_n}$ are constants which
are invariant under any permutation of the indices $\mu_1, \ldots,
\mu_n$, and $Q_{n}^{\mathcal{O}| \mu_1 \cdots \mu_n}(y_1,
\ldots,y_n)$ are entire functions of $y_i=2\cosh\theta_i$ with
$i=1,\ldots,n$. The factors in the denominator encode the full
pole structure of the form factors. More precisely, the kinematic
poles are encoded in the $(y_i+y_j)$ factor, whereas the functions
$P_{ab}(\theta)$ account for the bound state poles in the theory.
The form of these functions was determined in \cite{oota} for the
bulk case.  They are exactly the same in the boundary theory and
can be written as,
\begin{equation}
P_{ab}(\theta)=P_{\bar{a}\bar{b}}(\theta)=\prod_{\begin{tabular}{c}
   \scriptsize{ $p=|a-b|+1, p \neq h-1$} \\
    \text{\scriptsize step 2} \\
  \end{tabular}}^{a+b-1} 2\left[\cosh\theta- \cos\left(\frac{\pi
  (p+1)}{h}\right)\right], \label{pab}
\end{equation}
which is equivalent to Oota's expression (up to constant factors).
The value $p=h-1$ needs to be excluded as it would produce extra
poles at $\theta=i\pi$. The main difference in the way of encoding
the bound state pole structure with respect to the bulk theory is
the presence of products
$P_{ab}(\theta_{ab})P_{ab}(\hat{\theta}_{ab})$ rather than a
single factor $P_{ab}(\theta_{ab})$. This must be so in order to
guarantee invariance under change of sign of one of the
rapidities. Such prescription was already used in \cite{BPT} for
the Yang-Lee theory for which (\ref{pab}) is simply $2 \cosh
\theta +1$ (corresponding to $a=b=1$ and $h=3$) and the same holds
for the $A_{2}$-ATFT.

The ansatz (\ref{ansatz}) satisfies  all form factor consistency
equations by construction, provided that $Q_{n}(y_1,\ldots,y_n)$
has the extra property
\begin{equation}\label{Q}
  Q_{n}^{\mathcal{O}| \ldots \mu_a \mu_{a+1}
  \ldots}(\ldots,y_a,y_{a+1},\ldots)=Q_{n}^{\mathcal{O}| \ldots \mu_{a+1} \mu_{a}
  \ldots}(\ldots,y_{a+1},y_a,\ldots).
\end{equation}
From this property we deduce that $Q_{n}^{\mathcal{O}| \mu_1
\cdots \mu_n}(y_1, \ldots,y_n)$ must be a symmetric function of
all variables related to the same particle type. Therefore we
expect that the corresponding form factors can be expressed in
terms of elementary symmetric polynomials on the variables related
to each particle type. Plugging the ansatz (\ref{ansatz}) into the
kinematic residue equation we obtain recursive relations for the
polynomials $Q_{n}^{\mathcal{O}| \mu_1 \cdots \mu_n}(y_1,
\ldots,y_n)$ of the form
\begin{equation}\label{Qn}
Q_{n+2}^{\mathcal{O}| \bar{\mu} \mu \mu_1\cdots
\mu_n}(-y,y,y_1,\ldots,y_n)=P_n(y,y_1,\ldots,y_n)
Q_{n}^{\mathcal{O}| \mu_1 \cdots \mu_n}(y_1,\ldots,y_n),
\end{equation}
where $y=2 \cosh \theta$ and $P_{n}(y, y_1, \ldots, y_n)$ is a
polynomial on the variables $y,y_1,\ldots,y_n$. In addition, if
two particles $a,b$ fuse to produce a particle ${c}$, then also a
second set of equations of the form
\begin{eqnarray}
&& Q_{n+2}^{\mathcal{O}|a b \mu_1 \ldots
\mu_n}(y_{+},y_{-},y_1,\ldots,y_n) =W_{n}(y,y_1,\ldots,y_n)
Q_{n+1}^{\mathcal{O}|c \mu_1 \cdots
\mu_n}(y,y_1,\ldots,y_n),\label{Qnbound}
\end{eqnarray}
must be satisfied, where $y_{\pm}=2 \cosh\left(\theta\pm\frac{ i
u_{ab}^{c}}{2}\right)$ and $W_{n+1}(y, y_1, \ldots, y_n)$ is
 a polynomial of its variables. We will now proceed to compute the
 polynomials $P_n$ and $W_n$ for the simplest model of this class: the
 $A_{2}$-ATFT.
\section{Boundary form factors for $A_{2}$-ATFT} \label{s3}

\noindent We will now write down the equations
(\ref{Qn})-(\ref{Qnbound}) for the particular case $h=3$ and
$n=2$. We have then a two-particle theory with $\bar{1}=2$ and the
following fusing processes
\begin{eqnarray}
  && 1+1 \rightarrow 2 \quad \text{with} \quad u_{11}^{2}=\frac{2\pi}{3},\\
  && 2+2 \rightarrow 1 \quad \text{with} \quad u_{22}^{1}=\frac{2\pi}{3}.
\end{eqnarray}
Before writing down the recursive equations for this case, it is
convenient to introduce some notation. We will write
$F_{m+n}^{\mathcal{O}|m,n}(\{y\}_m;\{y'\}_n)$ to denote an
$n+m$-particle form factor where $m$ particles are of type 1 and
$n$ particles are of type 2. We choose to order indices in such a
way that all particles of type 1 appear first and all  particles
of type 2 appear at the end. In addition, we will group all
variables related to particles of type 1 or particles of type 2
into sets $\{y\}_m=\{y_1, \ldots, y_m\}$ and $\{y'\}_n=\{y'_1,
\ldots, y'_n\}=\{y_{m+1}, \ldots, y_{n+m}\}$, respectively.
Similar notation will  be used for the polynomials $Q$. Employing
this notation, we can rewrite equation (\ref{Qn}) as follows
\begin{equation}\label{Qn2}
Q_{m+n+2}^{\mathcal{O}|m+1,n+1}(y,\{y\}_m ;-y,
\{y'\}_n)=P_{m+n}(y,\{y\}_m ,\{y'\}_n)
Q_{m+n}^{\mathcal{O}|m,n}(\{y\}_m ;\{y'\}_n),
\end{equation}
where
\begin{equation}\label{Pn}
    P_{n+m}(y,\{y\}_m,\{y'\}_{n})=a(y)
    \sum_{k,p,r=0}^{m}\sum_{a,b,c=0}^{n}\left(\frac{\omega_{-}^{m-k}
\eta_{+}^{m-p} \lambda_{+}^{m-r}}{\tilde{\omega}_{-}^{a-n}
\tilde{\eta}_{+}^{b-n}
    \tilde{\lambda}_{+}^{c-n}}-
    \frac{\omega_{+}^{m-k} \eta_{-}^{m-p} \lambda_{-}^{m-r}}{\tilde{\omega}_{+}^{a-n} \tilde{\eta}_{-}^{b-n} \tilde{\lambda}_{-}^{c-n}}
    \right)\sigma_{r}\sigma_{p}\sigma_{k}\hat{\sigma}_{a}\hat{\sigma}_{b}\hat{\sigma}_{c},
\end{equation}
with
\begin{eqnarray}\label{a}
    a(y)&=&\frac{\cosh(2\theta)-\cos\left(\frac{\pi(2-B)}{3}\right)}{i\sqrt{3} \sinh
    \theta}, \\
    \omega_{\pm}&=&-2\cosh\left(\theta\pm \frac{2 \pi
    i}{3}\right), \quad \tilde{\omega}_{\pm}=-2\cosh\left(\theta\pm \frac{\pi
    i}{3}\right),\\
  \eta_{\pm}&=&-2\cosh\left(\theta\pm \frac{i\pi
    (2-B)}{3}\right), \quad \lambda_{\pm}=-2\cosh\left(\theta\pm \frac{i\pi
    B}{3}\right),\\
  \tilde{\eta}_{\pm}&=&-2\cosh\left(\theta\pm \frac{i\pi
    (3-B)}{3}\right), \quad \tilde{\lambda}_{\pm}=-2\cosh\left(\theta\pm \frac{i\pi
    (1+B)}{3}\right),\label{cons}
\end{eqnarray}
and $\sigma$ and $\hat{\sigma}$ are elementary symmetric
polynomials on the variables $\{y\}_m$ and $\{y'\}_n$
respectively. These polynomials can be defined by means of the
generating function,
\begin{equation}\label{sig}
    \prod_{k=1}^{n}(x+y_i)=\sum_{k=0}^{n}x^{n-k} \sigma_{k},
\end{equation}
where the subscript $k$ is the degree of the polynomial $\sigma_k$
which depends on the $n$ variables $y_1, \ldots, y_n$. Also from
the kinematic residue equations it follows that
\begin{equation}
    H_{m+n+2}^{\mathcal{O}|m+1,n+1}=\sqrt{3}f_{21}(i \pi)^{-1}
    H_{m+n}^{\mathcal{O}|m,n}.\label{cons1}
\end{equation}
In addition, due to the presence of bound states there is an
additional set of recursive equations of the form (\ref{Qnbound}).
Employing our new notation, we will write these equations as
\begin{eqnarray}
Q_{m+n+2}^{\mathcal{O}|m+2,n}(y_{+},y_{-},\{y\}_m ; \{y'\}_n)
=W_{m+n}(y,\{y\}_m,\{y'\}_n )
Q_{m+n+1}^{\mathcal{O}|m,n+1}(\{y\}_m; y ,
\{y'\}_n),\label{Qnbound2}
\end{eqnarray}
where
\begin{equation}\label{R}
W_{m+n}(y,\{y\}_m,\{y'\}_n )= g(B) \left(y^2-2-2
\cos\left(\frac{\pi B}{3}\right)\right)
    \sum_{k,r,s=0}^{m}y^{m-k}\tau_{+}^{m-r}\tau_{-}^{m-s} \sigma_{k}\sigma_{r}\sigma_{s},
\end{equation}
with
\begin{equation}\label{tau}
    \tau_{\pm}=-2\cosh\left(\theta \pm \frac{i
    \pi(1-B)}{3}\right),\qquad
    g(B)={\frac{2\Gamma_{11}^{2}}{3^{1/4}}\,
    \sin\left(\frac{\pi(2+B)}{6}\right)}.
\end{equation}
Notice that $g(1)=1$ and that $W_{m+n}$ does only depend on the
variables related to the type 1 particles. This surprising
property is due to (\ref{also}) and to the following identities
\begin{equation}
    \frac{f_{11}(2 \theta )w_{1}(\theta+ \frac{i \pi}{3})w_{1}(\theta- \frac{i
    \pi}{3})}{w_{1}(\theta)P_{11}(2\theta)}=\frac{1}{y^2-2-2 \cos\left(\frac{B \pi}{3}\right)},
\end{equation}
\begin{equation}
    \frac{f_{12}(\theta +\frac{i \pi}{3} )f_{12}(\theta- \frac{i
    \pi}{3})}{f_{22}(\theta)}=1, \qquad\qquad
    P_{12}(\theta)=1,\label{in}
\end{equation}
which hold specifically for the  $h=3$ case.  The bound state
residue equation provides also the following recursive relations
between the constants:
\begin{equation}
    H_{n+m+2}^{\mathcal{O}|m+2,n}=-\sqrt{\frac{\sqrt{3}}{f_{21}(i \pi)}}H_{m+n+1}^{\mathcal{O}|m,n+1},\label{cons2}
\end{equation}
were we have used  the identity
\begin{equation}
    f_{21}(i \pi)=\frac{4 f_{11}(2 \pi i/3)^2}{3}
    \sin^2\left(\frac{\pi(2+B)}{6}\right),
\end{equation}
which is valid for $h=3$. Equations (\ref{cons1})-(\ref{cons2})
are solved by
\begin{eqnarray}
H_{2n+3k}^{\mathcal{O}|n,n+3k}&=&(-1)^k
\left[\frac{\sqrt{3}}{f_{21}(i\pi)}\right]^{n+3k/2}
    H_{0}^{\mathcal{O}|0,0},\\
   H_{2n+3k+1}^{\mathcal{O}|n,n+1+3k}&=&(-1)^k \left[\frac{\sqrt{3}}{f_{21}(i\pi)}\right]^{n+3k/2}
    H_{0}^{\mathcal{O}|0,1},\\
   H_{2n+3k+2}^{\mathcal{O}|n,n+2+3k}&=&(-1)^{k+1} \left[\frac{\sqrt{3}}{f_{21}(i\pi)}\right]^{n+(3k+1)/2}
    H_{0}^{\mathcal{O}|0,1},
\end{eqnarray}
for $k \in \mathbb{Z}^{+}  \cup \{0\}$.

\subsection{Solving the form factor recursive equations}\indent\\

\noindent Finding solutions to (\ref{Qn2})-(\ref{Qnbound2}) is
very involved, even for the lower particle form factors. For this
reason here we have made a further simplification and consider
only the $B=1$ case. We will start by introducing some useful
definitions. We define the order of the form factor
$F_{m+n}^{\mathcal{O}|m,n}$ as the number
$[F_{m+n}^{\mathcal{O}|m,n}]$ which characterizes the asymptotic
behaviour
\begin{equation}
    \lim_{s \rightarrow \infty} F_{m+n}^{\mathcal{O}|m,n}(\{\theta+s\}_m; \{\theta+s\}_n)\sim e^{s[F_{m+n}^{\mathcal{O}|m,n}]}, \label{order}
\end{equation}
where by $\theta+s$ we mean that each rapidity is shifted by the
same amount $s$. If we were dealing with bulk form factors, this
would be simply the spin of the operator under investigation. The
order of the polynomial $Q_{m+n}^{\mathcal{O}|m,n}$ can be defined
in a similar fashion.

It will also be useful to introduce some particular combinations
of elementary symmetric polynomials, which we will denote by
$K^{[m,0]}$ and $Z^{[m,n]}$ and which are defined by the following
properties:
\begin{eqnarray}
  Z^{[m+1,n+1]}(y , \{y\}_{m}; -y, \{y'\}_{n}) = 0,\qquad  K^{[m+2,0]}(y_{+},y_{-}, \{y\}_{m}) =
  0,\label{kz}
\end{eqnarray}
namely, they are the kernels of the equations
(\ref{Qn2})-(\ref{Qnbound2}) for $B=1$. As it turns out, the
$K^{[1,n]}$ and $K^{[n,1]}$ polynomials admit a simple closed form
\begin{equation}\label{Z1n}
    Z^{[1,n]}=\sum_{k=0}^{n} \hat{\sigma}_k {\sigma}_1^{n-k}, \qquad Z^{[n,1]}=\sum_{k=0}^{n} {\sigma}_k \hat{\sigma}_1^{n-k},
\end{equation}
and each of the terms in the sums has order $n$. Here we will also
need
\begin{equation}
Z^{[2,2]}=\hat{\sigma}_2^2 +
{\hat{\sigma}_1}{\hat{\sigma}_2}{{\sigma }_1} +
{\hat{\sigma}_2}{\sigma }_1^2 + \hat{\sigma}_1^2{{\sigma }_2} -
2{\hat{\sigma}_2}{{\sigma }_2} +
  {\hat{\sigma}_1}{{\sigma }_1}{{\sigma }_2} + {\sigma }_2^2.
\end{equation}
These polynomials are in fact the same introduced in \cite{oota}
and denoted there by $K^{[m,n]}$. The $K$-polynomials defined by
the property (\ref{kz}) become soon rather complex and no obvious
pattern in terms of elementary symmetric polynomials seems to
emerge. For example:
\begin{eqnarray}
  K^{[2,0]} &=& 3-\sigma_1^2+\sigma_2, \\
  K^{[3,0]} &=& \sigma_1^3 (3 \sigma_1+\sigma_3)-\sigma_1^2 (3+ \sigma_2)(6+
    \sigma_2) + (3+ \sigma_2)^3,\\
  K^{[4,0]}&=&  -3K^{[2,0]}{( 3{{\sigma} }_1^2 - {( 3 +
{{{\sigma} }_2} ) }^2 ) }^2 + K^{[3,0]}
  {{{\sigma} }_3}( 9{{{\sigma} }_1} + {{{\sigma} }_3} )\nonumber \\
&&-3
  {{{\sigma} }_3}( (\sigma_1((3+\sigma_2)^2-3 \sigma_1^2)-4 \sigma_3
  (3+\sigma_2))( 3 + {{{\sigma} }_2} )  +
        ( {{\sigma} }_1^4 + 4{( 3 + {{{\sigma} }_2} ) }^2 ) {{{\sigma} }_3} )
        \nonumber\\
        && +
  {{{\sigma} }_4}( -27{{\sigma} }_1^4 + ( 3 - {{{\sigma} }_2} ) {( 3 + {{{\sigma} }_2} ) }^3 +
     {{\sigma} }_1^2( 3 + {{{\sigma} }_2} ) ( 27 + {{{\sigma} }_2}( 6 + {{{\sigma} }_2} )  )\\
     &&  -
     {{{\sigma} }_1}{{{\sigma} }_3}( -9 + {{\sigma} }_2^2 + 3{{{\sigma} }_1}( 6{{{\sigma} }_1} + {{{\sigma} }_3} )  )
         + ( (3+{{{\sigma} }_2}) (3+2{{{\sigma} }_2}) +
        3{{{\sigma} }_1}( 3{{{\sigma} }_1} + {{{\sigma} }_3} )  - {{{\sigma} }_4} ){{{\sigma} }_4}
        ).\nonumber
\end{eqnarray}
Similarly, we define the $K^{[0,n]}$ polynomials as identical to
the ones above up to the replacement $\sigma \rightarrow
\hat{\sigma}$. Like those, they have order $n(n-1)$. In this case,
not all terms involved are of order $n(n-1)$ but only the leading
ones. In fact if we select out only the leading terms we get a new
set of polynomials which will be related to structures occurring
in the bulk form factors:
\begin{eqnarray}
  K^{[2,0]}_{\text{bulk}} &=&\sigma_2 -\sigma_1^2, \\
  K^{[3,0]}_{\text{bulk}} &=& \sigma_1^3 \sigma_3+ \sigma_2^2   K^{[2,0]}_{\text{bulk}},\\
  K^{[4,0]}_{\text{bulk}}&=&  \sigma_3^2
  K^{[3,0]}_{\text{bulk}}-\sigma_2^3 \sigma_4
  K^{[2,0]}_{\text{bulk}}-\sigma_4 \sigma_1 \sigma_3(\sigma_2^2+3\sigma_1 \sigma_3-3
  \sigma_4)+ \sigma_4^2 (2\sigma_2^2-\sigma_4).
\end{eqnarray}
These ``bulk" polynomials are in fact equivalent (up to a sign) to
the polynomials $B_{1[m,n]}$ introduced in \cite{oota} and can be
expressed as determinants of elementary symmetric polynomials in
the way described there. We have now all the tools to find
explicit solutions to the form factor equations.

\subsection{Form factors of ``spinless" fields}\indent \\

\noindent In this subsection we will compute all form factors up
to four particles of a family of fields, which we will denote by
$\mathcal{O}_2$ and which we will try to identify later. In the
title of this subsection we have used the word ``spinless" to
indicate that the order of the form factors associated to these
fields is zero and therefore they should correspond to spinless
fields in the bulk theory. Indeed, it will turn out that this
family of operators is nothing but the boundary counterpart of the
fields whose form factors were obtained by T.~Oota \cite{oota}.
The condition $[F_{m+n}^{\mathcal{O}_2|m,n}]=0$ fixes the order of
each polynomial $Q_{m+n}^{\mathcal{O}|m,n}$ as
\begin{equation}
[Q_{m+n}^{\mathcal{O}|m,n}]=m^2+ (m+n)(n-1).
\end{equation}
\subsubsection{Vacuum expectation values and 1-particle form
factors} Here we will normalize $Q_{0}^{\mathcal{O}_2|0,0}=1$ and
choose
\begin{equation}
    Q_{1}^{\mathcal{O}_2|0,1}=A_{[0,1]},\label{q1}
\end{equation}
with $A_{[0,1]}$ an arbitrary constant.

\subsubsection{Two-particle form factors}
The recursive equations for the two-particle form factors are:
\begin{eqnarray}
    Q_{2}^{\mathcal{O}_2|1,1}(-y;y)&=& 0,\\
    Q_{2}^{\mathcal{O}_2|2,0}(y_{+},y_{-})&=&(y^2-3)A_{[0,1]}.
\end{eqnarray}
Both equations admit many solutions (in fact, the first equation
admits infinitely many). We will start by selecting those
solutions whose order (\ref{order}) is minimal. We then obtain
\begin{eqnarray}
    Q_{2}^{\mathcal{O}_2|1,1}(y_1;y'_1)&=& \sigma_1 +\hat{\sigma}_1 ,\label{11}\\
    Q_{2}^{\mathcal{O}_2|2,0}(y_1,y_2)&=&A_{[0,1]}(\sigma_1^2-3)+ A_{[2,0]}K^{[2,0]},
\end{eqnarray}
with $A_{[2,0]}$ constant.
\subsubsection{Three-particle form factors}
There are three different three-particle form factors that can be
obtained from the solutions above: $Q_3^{\mathcal{O}_2|1,2}$,
$Q_3^{\mathcal{O}_2|3,0}$ and $Q_3^{\mathcal{O}_2|0,3}$. The first
one is the solution of the equations
\begin{eqnarray}
    Q_{3}^{\mathcal{O}_2|1,2}(y;-y,y_1)&=& (y^2-3)(3+y y_1-y_1^2) A_{[0,1]}, \\
    Q_3^{\mathcal{O}_2|1,2}(y_{1};y_+,y_{-})&=&(y^2-3)(
    A_{[0,1]}((y+y_1)^2-3)+ A_{[2,0]}(3-y^2-y_1^2- y y_1)).
\end{eqnarray}
The most general solution of these equations with the required
order is
\begin{eqnarray}
  Q_{3}^{\mathcal{O}_2|1,2}(y_1;y'_1,y'_2) = A_{[1,2]}Z^{[1,2]}K^{[0,2]}-\hat{\sigma}_2 (A_{[2,0]}-A_{[0,1]}) Z^{[1,2]}+ A_{[0,1]}(
   \hat{\sigma}_2\hat{ \sigma}_1 \sigma_1-3 K^{[0,2]}),
\end{eqnarray}
which once again involves a new arbitrary constant $A_{[1,2]}$.
Let us consider now the three-particle form factor involving only
type 1 particles. The equation we need to solve is:
\begin{equation}
    Q_{3}^{\mathcal{O}_2|3,0}(y_{+},y_{-},y_1)=(y^2-3)(y^2-y_1^2)^2,
\end{equation}
and the most general solution reads
\begin{eqnarray}
    Q_{3}^{\mathcal{O}_2|3,0}(\{y\}_3)=\sigma_1 J^{[3,0]} + A_{[3,0]}
    K^{[3,0]},
\end{eqnarray}
with
\begin{eqnarray}
  J^{[3,0]}&=&\sigma_1((3+\sigma_2)^2-3 \sigma_1^2)-4 \sigma_3
  (3+\sigma_2),
\end{eqnarray}
and $A_{[3,0]}$ an arbitrary constant.  Finally, since the form
factor (\ref{11}) is symmetric in its variables,
$Q_{3}^{\mathcal{O}_2|0,3}$ is given by exactly the same function
as $Q_{3}^{\mathcal{O}_2|3,0}$.
\subsubsection{Four-particle form factors}
There are three four-particle form factors which are related to
the solutions above by the boundary and kinematic residue
equations: $Q_4^{\mathcal{O}_2|3,1}$, $Q_4^{\mathcal{O}_2|0,4}$
and $Q_4^{\mathcal{O}_2|2,2}$. The polynomial
$Q_4^{\mathcal{O}_2|3,1}$ is the solution to
\begin{eqnarray}
    Q_{4}^{\mathcal{O}_2|3,1}(y_1,y_2,-y;y)&=& Q_{2}^{\mathcal{O}_2|2,0}(y_1,y_2) P_{0+2}(y,y_1,y_2),\\
    Q_4^{\mathcal{O}_2|3,1}(y_{+},y_{-},y_1;y_2)&=&Q_{3}^{\mathcal{O}_2|1,2}(y_1;y_2,y)W_{1+1}(y,y_1),
\end{eqnarray}
which is given by
\begin{eqnarray}
&& Q_{4}^{\mathcal{O}_2|3,1}(\{y\}_3;y'_1) =(A_{[0,1]}-A_{[2,0]})
  [{{\sigma }_1}{{\sigma }_3}( {\sigma }_1^2( -9 + {{\sigma }_2} )  + 12( 3 + {{\sigma }_2} )  -
       4{{\sigma }_1}{{\sigma }_3} ) \hat{\sigma}_1 \nonumber \\
       &&\qquad + J^{[3,0]}( 3 + {{\sigma }_2} ) \hat{\sigma}_1^2
       -
  3K^{[3,0]}( {{\sigma }_1} + 4\hat{\sigma}_1 )  ] -
  A_{[0,1]}(3Z^{[2,1]}{{\sigma }_1}( 3{\sigma }_1^2 - {( 3 + {{\sigma }_2} ) }^2 ) \nonumber \\
  && \qquad+
  3{{\sigma }_3}( 4{( 3 + {{\sigma }_2} ) }^2 + {\sigma }_1^3 \hat{\sigma}_1 )  + {{\sigma }_3}
  ( {{\sigma }_1}( 9 - {{\sigma }_2} )  + 4{{\sigma }_3} )
   ( 3(  \hat{\sigma}_1 -{{\sigma }_1})  +\hat{\sigma}_1( {{\sigma }_2} + {{\sigma }_1}\hat{\sigma}_1 )
   ))\nonumber \\
 & & \qquad-  (A_{[3,1]}K^{[3,0]}+ A_{[1,2]}(K^{[3,0]}+\sigma_1
 J^{[3,0]}))Z^{[3,1]},
\end{eqnarray}
where $A_{[3,1]}$ is a new arbitrary constant. Let us now compute
\begin{eqnarray}
    Q_4^{\mathcal{O}_2|0,4}(y_+,y_-,y_{1},y_{2})&=&Q_{3}^{\mathcal{O}_2|1,2}(y;y_1,y_2)W_{2+0}(y,y_1,y_2),
\end{eqnarray}
which is solved by,
\begin{eqnarray}
   Q^{\mathcal{O}_2|0,4}(\{y'\}_4) &=& A_{[0,4]}K^{[0,4]} -A_{[2,0]}\hat{\sigma}_4(-6 \hat{\sigma}_1^2 + (3 + \hat{\sigma}_2)(9 + \hat{\sigma}_2) - 2 \hat{\sigma}_1 \hat{\sigma}_3
   +2 \hat{\sigma}_4)^2 \nonumber \\
  &+& A_{[0,1]}[3( {{{\hat{\sigma} }_1}}^2( -9 + {{\hat{\sigma} }_2} )  + 12( 3 + {{\hat{\sigma} }_2} )  -
     4{{\hat{\sigma} }_1}{{\hat{\sigma} }_3} ) (
     {( 3{{\hat{\sigma} }_1} + {{\hat{\sigma} }_3} ) }^2 -3{( 3 + {{\hat{\sigma} }_2} ) }^2 ) \nonumber \\
     &&\qquad +
  {{\hat{\sigma} }_4}( 36{{{\hat{\sigma} }_1}}^4 + {( 3 + {{\hat{\sigma} }_2} ) }^2
      ( -45 + {{\hat{\sigma} }_2}( 24 + {{\hat{\sigma} }_2} )  )  -
     3{{{\hat{\sigma} }_1}}^2( 9 + {{\hat{\sigma} }_2}( 36 + 7{{\hat{\sigma} }_2} )  )  \nonumber \\
     && \qquad+
     {{\hat{\sigma} }_3}( 24{{{\hat{\sigma} }_1}}^3 - 3{{\hat{\sigma} }_1}( -1 + {{\hat{\sigma} }_2} ) ( 9 + {{\hat{\sigma} }_2} )  +
        4{{\hat{\sigma} }_3}({{{\hat{\sigma} }_1}}^2 - {{\hat{\sigma} }_2}) ) \nonumber \\
     && \qquad +
     {{{\hat{\sigma}} }_4}( {{{\hat{\sigma}} }_2}( -3 - 4{{{\hat{{\sigma}} }_1}}^2 + 7{{\hat{{\sigma}} }_2} )  + 4{{\hat{{\sigma}} }_1}
     {{{\hat{\sigma}} }_3} -
        8( 9 + {{{\hat{\sigma}} }_4} )  ))]\nonumber \\
        & + & A_{[1,2]}[-3( 3 + {{\hat{\sigma} }_2} ) ( {{{\hat{\sigma} }_1}}^2( -9 + {{\hat{\sigma} }_2} )  +
     12( 3 + {{\hat{\sigma} }_2} )  ) ( 3{{{\hat{\sigma} }_1}}^2 - {( 3 + {{\hat{\sigma} }_2} ) }^2 )\nonumber \\
     && \qquad -
   {{\hat{\sigma} }_3}( 3 + {{\hat{\sigma} }_2} )( 6{{{\hat{\sigma} }_1}}^3( -15 + {{\hat{\sigma} }_2} )  +
     {{{\hat{\sigma} }_1}}^2( -33 + {{\hat{\sigma} }_2} ) {{\hat{\sigma} }_3} + 12( 3 + {{\hat{\sigma} }_2} ) {{\hat{\sigma} }_3})\nonumber \\&&\qquad
     -4{{\hat{\sigma} }_1}\hat{\sigma}_3( 3 + {{\hat{\sigma} }_2} )
     (3(3+{{\hat{\sigma} }_2})(9+{{\hat{\sigma} }_2}) - {{{\hat{\sigma} }_3}}^2 ) +
     12{{{\hat{\sigma} }_1}}^4 {{\hat{\sigma} }_4} ( -3 + {{\hat{\sigma} }_2} ) \nonumber \\ &&
    \qquad +{{\hat{\sigma} }_4}( 3 + {{\hat{\sigma} }_2} )( -
    6{{{\hat{\sigma} }_1}}^2 ( 3 + 5{{\hat{\sigma} }_2} )  +
    {( 3 + {{\hat{\sigma} }_2} ) }( 81 + 42{{\hat{\sigma} }_2} + {{{\hat{\sigma} }_2}}^2 ))\nonumber \\ &&
   \qquad  -
    2{{\hat{\sigma} }_3} {{\hat{\sigma} }_4}( 2{{{\hat{\sigma} }_1}}^3( 9 - {{\hat{\sigma} }_2} )  +
       3{{\hat{\sigma} }_1}( 1 + {{\hat{\sigma} }_2} ) ( 3 + {{\hat{\sigma} }_2} ) +
       2( 3 + {{\hat{\sigma} }_2} + 2 {{{\hat{\sigma} }_1}}^2 ) {{\hat{\sigma} }_3} ) \nonumber \\&&
       \qquad +  {{\hat{\sigma}
       }_4^2}( 4{{{\hat{\sigma} }_1}}^2( 9 - {{\hat{\sigma} }_2} )  +
       ( 3 + {{\hat{\sigma} }_2} ) ( -15 + 7{{\hat{\sigma} }_2} )  + 8(2{{\hat{\sigma} }_1}{{\hat{\sigma} }_3} - {{\hat{\sigma}
       }_4})
       )],
\end{eqnarray}
where $A_{[0,4]}$ is another arbitrary constant. Finally, we need
to solve for the polynomial $Q^{\mathcal{O}_2|2,2}_4$. It is the
solution of the following three equations:
\begin{eqnarray}
  Q_4^{\mathcal{O}_2|2,2}(y_+,y_-;y_1,y_2) &=& (y^2-3)Q_3^{\mathcal{O}_2|0,3}(y,y_1,y_2), \\
  Q_{4}^{\mathcal{O}_2|2,2}(y_1,y_2;y_+,y_-) &=&
  (y^2-3)Q_3^{\mathcal{O}_2|3,0}(y,y_1,y_2),\\
  Q_{4}^{\mathcal{O}_2|2,2}(y,y_1;-y,y_2)&=&
  P_{1+1}(y,y_1,y_2)Q^{\mathcal{O}_2|1,1}_2(y_1;y_2).
\end{eqnarray}
Since the polynomials $Q_3^{\mathcal{O}_2|0,3}$ and
$Q_3^{\mathcal{O}_2|3,0}$ are identical, the problem is reduced to
solving the last two equations and imposing invariance under the
transformation $\sigma\leftrightarrow \hat{ \sigma}$. The result
is
\begin{eqnarray}
Q_{4}^{\mathcal{O}_2|2,2}(\{y\}_2;\{y'\}_2)&=& 9{\hat{\sigma}_2}(
3{\sigma }_1^2 - {\sigma }_1^4 - 2{{\sigma }_2} )  -
  3{{\sigma }_1}{{\sigma }_2}( -6{\hat{\sigma}_1} + 2{\hat{\sigma}_1}{\sigma }_1^2 + {{\sigma }_1}{{\sigma }_2} )\nonumber \\
  &+&
  \hat{\sigma}_1^3( {\hat{\sigma}_1} + 2{{\sigma }_1} ) ( -3K^{[2,0]} + {\sigma }_2^2 )  +
  3K^{[2,0]}( 3{{\sigma }_1}( 2{\hat{\sigma}_1} + {{\sigma }_1} )  + {\sigma }_2^2 )\nonumber \\
  &+&
  {{\hat{\sigma}_1}}^2( 3(K^{[2,0]})^2 + 3K^{[2,0]}{{\sigma }_2} + {\sigma }_2^3 )  +
  {{\hat{\sigma}_2}}^2( 9 + {\hat{\sigma}_1}{{{\sigma }_1}}^3 + ( 3 + \hat{\sigma}_1^2 ) {{\sigma }_2}
  )\nonumber \\
&-&
  3 {{\sigma }_1} {{\hat{\sigma}_2}}^2( 1 + {{\sigma }_2} )(
    {\hat{\sigma}_1} + {{{\sigma }_1}})
   +
  {\hat{\sigma}_2}( 3( 3{\sigma }_1^2 - 2{{\sigma }_2} ) {{\sigma }_2}
  -
     {\hat{\sigma}_1}{\sigma }_1^3( 12 - {{\sigma }_2} )
     )\nonumber \\
 &+&
  {\hat{\sigma}_2}(
        3( 4 - {{\sigma }_2} ) ( 3 + {{\sigma }_2}   )  +
     {{\hat{\sigma}_1}}^2( 9 - {\sigma }_1^2( 3 - 5{{\sigma }_2} )  -
        {{\sigma }_2}( 6 + 5{{\sigma }_2}- 2{\hat{\sigma}_1}{{\sigma }_1} )  )  )
\nonumber\\
 &+ & K^{[2,0]}Z^{[2,2]}( {A_{[2,2]}}K^{[2,0]} -3(2 -
A_{[3,0]}) - {\hat{\sigma}_2}(1- A_{[3,0]})
       ) \nonumber \\ &-& Z^{[2,2]}A_{[3,0]}( 9 - 3{\sigma }_1^2 + \hat{\sigma}_1^2{{\sigma }_2} - {\hat{\sigma}_2}{{\sigma
       }_2}),
\end{eqnarray}
with $A_{[2,2]}$ constant.
\subsubsection{Remarks on operator identification}
In the previous subsection, all our form factors have ``descended"
either from $Q_{1}^{\mathcal{O}_2|0,1}$ or from
$Q_{0}^{\mathcal{O}_2|0,0}$, so that we had no need to fix
$Q_{1}^{\mathcal{O}_2|1,0}$. One obvious option is to choose
$Q_{1}^{\mathcal{O}_2|0,1}=Q_{1}^{\mathcal{O}_2|1,0}$ and, more
generally, to consider form factor solutions which are completely
symmetric under the exchange of the names of the particles. Such
form factors should correspond to a particular type of operators
which we will call ``symmetric". We mean by this, that they should
be symmetric on the fundamental fields, such as
$\mathcal{O}_{12}=\phi_1 +\phi_2$ and
$\hat{\mathcal{O}}_{12}=\phi_1
\partial_x \phi_2 + \phi_2 \partial_x \phi_1$. But there are of course many fields
 which are not symmetric, for example those that depend only either on $\phi_1$
(``type-1 fields") or on $\phi_2$ (``type-2 fields"). It is
natural to assume that for any such fields the one particle form
factors $Q_{1}^{\mathcal{O}_2|0,1}$ (for type-1) or
$Q_{1}^{\mathcal{O}_2 |1,0}$ (for type 2) should vanish.
Therefore, the solutions which we just constructed could
correspond to a ``type-2 field" $\mathcal{O}_2$ if we impose the
condition that all form factors descending from
$Q_{1}^{\mathcal{O}_2|1,0}$ are vanishing. In order to identify
the precise operator our solutions correspond to we should analyze
the UV behaviour of the two-point function. We can however get
some idea by just analyzing the order of the form factors
obtained. As anticipated before
 \begin{equation}
[F_{m+n}^{\mathcal{O}_2|m,n}]=0.
\end{equation}
This indicates that the boundary field to which the solutions
correspond to is a spinless field in the bulk theory. If we
assume, as mentioned above, that this field has
$F_{1}^{\mathcal{O}_2|1,0}=0$, then a natural candidate is the
fundamental field $\phi_2$ and any powers thereof. This
constitutes an infinite countable set of fields, in correspondence
with the infinite countable set of solutions which the form factor
equations seem to have. We should now provide some more reasoning
to justify our last statement: notice that in all our examples
each new form factor involves a new arbitrary constant. If we
assume that this phenomenon will continue for higher particle form
factors, then the full set of solutions to our equations will
depend on a infinite but countable number of arbitrary parameters.
One way of understanding this is to assume that the solutions
obtained correspond in fact to a field which is a linear
combination of an infinite but countable set of ``type-2" fields,
in one-to-one correspondence with the constants $A_{[m,n]}$ (since
the form factor of a sum of fields equals the sum of their form
factors).

\subsubsection{From boundary to bulk form factors}
It is easy to see that from every boundary form factor solution, a
bulk form factor solution can be obtained by the simple procedure
of shifting all rapidities to infinity and selecting out the
leading order terms \cite{st}. A way to see this is to notice that
when such shift is performed, the boundary form factor consistency
equations tend to the bulk equations because of the properties
\begin{equation}
    \lim_{\theta \rightarrow \infty} R_{i}(\theta)=  \lim_{\theta \rightarrow \infty} w_{i}(\theta)=
    \lim_{\theta \rightarrow \infty} S_{ij}(\theta)= \lim_{\theta \rightarrow \infty}
    f_{ij}(\theta)=1.
\end{equation}
These imply that equations (\ref{eq2}) and (\ref{eq3}) become
equivalent to each other and equal to the standard crossing
relation for the bulk form factors. At the same time, the
$S$-matrix product depending on the sum of rapidities becomes 1 in
(\ref{eq4}) rendering it in its bulk form \cite{KW,smirnovbook}.

It is therefore interesting to consider the limit described above
for the solutions we have just obtained
\begin{eqnarray}
  && Q_{1; \text{bulk}}^{\mathcal{O}_2|0,1}(y'_1) = A_{[0,1]}, \\
 && Q_{2; \text{bulk}}^{\mathcal{O}_2|2,0}(\{y\}_2) = A_{[0,1]}\sigma_1^2 + A_{[2,0]}K^{[2,0]}_{\text{bulk}},\\
  && Q_{2; \text{bulk}}^{\mathcal{O}_2|1,1}(y_1;y'_1) = \sigma_1 + \hat{\sigma}_1,\\
  && Q_{3; \text{bulk}}^{\mathcal{O}_2|1,2}(y_1;y'_1, y'_2) =(A_{[1,2]}K^{[0,2]}_{\text{bulk}}
  -\hat{\sigma}_2 (A_{[2,0]}-A_{[0,1]})) Z^{[1,2]}+ A_{[0,1]}\hat{\sigma}_2\hat{ \sigma}_1 \sigma_1,\\
  &&Q_{3; \text{bulk}}^{\mathcal{O}_2|3,0} (\{y\}_3)= \sigma_1 \sigma_2(\sigma_1 \sigma_2-4 \sigma_3 ) +
  A_{[3,0]}K^{[3,0]}_{\text{bulk}},\\
  &&Q_{4; \text{bulk}}^{\mathcal{O}_2|3,1}(\{y\}_3;y'_1) =  \hat{\sigma}_1(A_{[0,1]}-A_{[2,0]})
 ({\sigma }_1^2{{\sigma }_3}+ {{\sigma }_2^2}\hat{\sigma}_1)(\sigma_1 \sigma_2-4 \sigma_3)
 -A_{[3,1]}K^{[3,0]}_{\text{bulk}}Z^{[3,1]}
 \nonumber \\
       && \qquad\qquad -A_{[1,2]}K^{[3,0]}_{\text{bulk}}Z^{[3,1]}+(\sigma_1 \sigma_2-4\sigma_3)(
  A_{[0,1]}  \hat{\sigma}_1{{\sigma }_3}
   ( {{\sigma }_2} + {{\sigma }_1}\hat{\sigma}_1 )-  A_{[1,2]}\sigma_1\sigma_2
   Z^{[3,1]}),\label{4p}\\
 && Q_{4; \text{bulk}}^{\mathcal{O}_2|0,4}(\{y'\}_4) =A_{[0,4]}K^{[0,4]}_{\text{bulk}} -A_{[2,0]}\hat{\sigma}_4( \hat{\sigma}_2^2 - 2 \hat{\sigma}_1 \hat{\sigma}_3
   +2 \hat{\sigma}_4)^2 \nonumber \\
  &&\qquad \qquad + A_{[0,1]}{{\sigma }_4}[ 4( {{\sigma }_1}{{\sigma }_3} - {{\sigma }_4} )
       ( {{\sigma }_1}{{\sigma }_3} + 2{{\sigma }_4} ) +
      {\sigma }_2^2( {\sigma }_2^2 - 3{{\sigma }_1}{{\sigma }_3} + 7{{\sigma }_4}
      )- 4{{\sigma }_2}( {\sigma }_3^2 + {\sigma }_1^2{{\sigma }_4} )   ]\nonumber \\ &
      & \qquad \qquad +
      A_{[1,2]}[-{{\sigma }_1}{{\sigma }_2}{\sigma }_3^2( {{\sigma }_1}{{\sigma }_2} - 4{{\sigma }_3} )    +
    {\sigma }_2^2{{\sigma }_4}({\sigma }_2^2 - 6{{\sigma }_1}{{\sigma }_3})+ (
     4{{\sigma }_2}( {\sigma }_1^3 - {{\sigma }_3} ) {{\sigma }_3}- 8{\sigma }_1^2{\sigma }_3^2 ) {{\sigma }_4}\nonumber \\ && \qquad\qquad
     + ( 7{\sigma }_2^2 - 4{{\sigma }_1}( {{\sigma
}_1}{{\sigma }_2} - 4{{\sigma }_3} )
     ) {\sigma }_4^2 - 8{\sigma }_4^3   ],\\
&& Q_{4;\text{bulk}}^{\mathcal{O}_2|2,2}(\{y\}_2;\{y'\}_2)=
{\hat{\sigma}_1}{\hat{\sigma}_2}{{\sigma }_2} ( 2
\hat{\sigma}_1^2{{\sigma }_1} + {{\sigma }_1}( {\sigma }_1^2 -
3{{\sigma }_2} )  -
     5{\hat{\sigma}_1}K^{[2,0]}_{\text{bulk}}    )  \nonumber \\ && \qquad\qquad +
 {\sigma }_2^2( \hat{\sigma}_1^2( {\hat{\sigma}_1}( {\hat{\sigma}_1} + 2{{\sigma }_1} )  + {{\sigma }_2} )
  +
  {\sigma }_1^2( {\hat{\sigma}_1}{{\sigma }_1} - 3{{\sigma }_2} ) +
     {\hat{\sigma}_1}( {\hat{\sigma}_1} - 3{{\sigma }_1} ) {{\sigma }_2}
     ) \nonumber \\
&& \qquad\qquad+ Z^{[2,2]}(( A_{[2,2]} K^{[0,2]}_{\text{bulk}}   -
{\hat{\sigma}_2} ) K^{[2,0]}_{\text{bulk}}  +
  {A_{[3,0]}}(\hat{ \sigma}_2 K^{[2,0]}_{\text{bulk}}+{ \sigma}_2 K^{[0,2]}_{\text{bulk}} )).
\end{eqnarray}
We can now compare these solutions to the ones obtained in
\cite{oota}. It turns out that, up to the three-particle form
factors, they completely agree with those when setting $q=1$
(which is equivalent to $B=1$ in our notation) and matching the
constants appropriately.  There is also agreement for the
four-particle form factor (\ref{4p}) but there are some important
differences between our solutions and those of T.~Oota for other
four-particle form factors, even though their orders agree. The
precise differences are:
\begin{eqnarray}
Q_{4;\text{bulk}}^{\mathcal{O}_2|0,4}(\{y'\}_4)-Q_{4;\text{Oota}}^{\mathcal{O}_2|0,4}(\{y'\}_4)=
  2{A_{[0,1]}}{{\sigma }_4^2}( {{\sigma }_1}{{\sigma }_3}-
     {\sigma }_4 ) +(A_{[0,1]}-
{A_{[2,0]}}){\sigma }_2^2{{\sigma }_4}( {\sigma }_2^2 + 2{{\sigma
}_4} ),
\end{eqnarray}
and
\begin{eqnarray}
&&
Q_{4;\text{bulk}}^{\mathcal{O}_2|2,2}(\{y\}_2;\{y'\}_2)-Q_{4;\text{Oota}}^{\mathcal{O}_2|2,2}(\{y\}_2;\{y'\}_2)=
3Z^{[2,2]}(1-{A_{[3,0]}} ) ( K^{[2,0]} - {{\hat{\sigma}_1}}^2 )
\nonumber \\ && \qquad +
  {{\hat{\sigma}_1}}^2{\hat{\sigma}_2}( {\sigma }_1^2 + {{\sigma }_2} ) ( 3 - K^{[2,0]} )  -
  {\hat{\sigma}_2}{{\sigma }_1}( {\hat{\sigma}_1}{{\sigma }_2}( {\hat{\sigma}_2} + {{\sigma }_2} )  +
     {{\sigma }_1}( {{\hat{\sigma}_1}}^2{\sigma }_1^2 + {\hat{\sigma}_2}{{\sigma }_2} )
     ).
\end{eqnarray}
We believe that Oota's solutions for these two cases must be
wrong, as the agreement of all other form factors with ours
strongly suggests that we are dealing with the same class of
fields.
\subsection{Form factors of ``spin-1" fields}\indent \\

\noindent In this section we want to find further solutions to the
boundary form factors equations and, as a by-product, also new
solutions to the bulk form factor equations. We will characterize
those solutions by the choice
\begin{equation}
    Q_1^{\hat{\mathcal{O}}_{2}|0,1}(y)=B_{[0,1]} y,\label{1p}
\end{equation}
where $B_{[0,1]}$ is an arbitrary constant. In addition, we would
like our form factors to have
\begin{equation}
[F_{m+n}^{\hat{\mathcal{O}}_2|m,n}]=1, \qquad\Leftrightarrow
\qquad [Q_{m+n}^{\hat{\mathcal{O}}_2|m,n}]=m^2+ (m+n)(n-1)+1,
\end{equation}
such that the field $\hat{\mathcal{O}}_2$ can be matched to a
spin-1 field of the bulk theory. If we choose
$Q_1^{{\hat{\mathcal{O}}_2|1,0}}=0$, we can identify this field as
being of ``type-2" again (like $\partial_x \phi_2$). An obvious
solution to these constraints is obtained by multiplying all
solutions of the previous subsection by the polynomial $\sigma_1 +
\hat{\sigma}_1$. If we do so we can identify
$\hat{\mathcal{O}}_2=\partial_x \mathcal{O}_2$, but other less
trivial solutions also exist, as we show below.

\subsubsection{Two-particle form factors}
A set of solutions, which is compatible with (\ref{1p}) is
\begin{eqnarray}
    Q_{2}^{\hat{\mathcal{O}}_{2}|1,1}(y_1,y'_1)&=& \hat{\sigma}_1 (\sigma_1 +\hat{\sigma}_1) ,\label{2p}\\
    Q_{2}^{\hat{\mathcal{O}}_{2}|2,0}(y_1,y_2)&=&  B_{[0,1]}\sigma_1 (\sigma_1^2-3)+(\hat{A}_{[2,0]} +B_{[2,0]}
    \sigma_1)K^{[2,0]},
\end{eqnarray}
where $\hat{A}_{[2,0]}$ and $B_{[2,0]}$ are constants.
\subsubsection{Three-particle form factors}
The three-particle form factors are given by
\begin{eqnarray}
  && Q_{3}^{\hat{\mathcal{O}}_{2}|1,2}(y_1,y'_1,y'_2) = (\hat{A}_{[1,2]}+(B_{[1,2]}-B_{[2,0]})\sigma_1+C_{[1,2]}\hat{\sigma}_1)Z^{[1,2]}K^{[0,2]}\nonumber \\
  &&\qquad -\hat{\sigma}_2 (\hat{A}_{[2,0]}+B_{[2,0]}(\sigma_1+\hat{\sigma}_1)) Z^{[1,2]}
  - B_{[0,1]}[3 K^{[0,2]}\hat{\sigma }_1 + \sigma_1^2 \hat{\sigma}_1( -6 + 2 \hat{\sigma}_1^2 - 5\hat{\sigma}_2
   ) \nonumber \\
  &&\qquad  +
  \sigma_1^3( -6 + 2\hat{\sigma}_1^2 - 3 \hat{\sigma}_2 )  - \hat{\sigma}_1\hat{\sigma}_2^2
  -
  3{\sigma_1}( -3 + \hat{\sigma}_1^2 + \hat{\sigma}_2+ \hat{\sigma}_2^2
  )],\\
&& Q_{3}^{\hat{\mathcal{O}}_{2}|3,0}(\{y\}_3)=(\sigma_2+3)
J^{[3,0]} + \sigma_1 \hat{A}_{[3,0]}
    K^{[3,0]},\\
&& Q_{3}^{\hat{\mathcal{O}}_{2}|0,3}(\{y'\}_3)= \hat{\sigma}_3(
\hat{\sigma}_1^2-2(3+\hat{\sigma}_2))^2 + (B_{[0,3]}+
C_{[0,3]}\hat{\sigma}_1) K^{[0,3]},
\end{eqnarray}
where all variables $\hat{A}_{[a,b]}$, $B_{[a,b]}$ and $C_{[a,b]}$
are constants. Notice that, since (\ref{2p}) is not anymore a
symmetric function, the polynomials
$Q_{3}^{\hat{\mathcal{O}}_{2}|3,0}$ and
$Q_{3}^{\hat{\mathcal{O}}_{2}|0,3}$ are now different from each
other.
\subsubsection{Four-particle form factors}
The four-particle form factors are given by
\begin{eqnarray}
&& Q_{4}^{\hat{\mathcal{O}}_{2}|3,1}(\{y\}_3,y'_1)= (\hat{A}_{[3,1]} + B_{[3,1]}\sigma_1 +C_{[3,1]}\hat{\sigma}_1)K^{[3,0]}Z^{[3,1]}\nonumber \\
 && \qquad- \hat{A}_{[1,2]}(( 3 + {{\sigma }_2} ) ( -3{\sigma }_1^2 + {( 3 + {{\sigma }_2} ) }^2 )  +
  {{\sigma }_1}( {\sigma }_1^2 - 4( 3 + {{\sigma }_2} )  ) {{\sigma }_3})Z^{[3,1]}\nonumber \\
 && \qquad- (B_{[1,2]}-B_{[2,0]})(-4 K^{[3,0]}{\hat{\sigma}_1} + {( {\sigma }_1^2 - 2( 3 + {{\sigma }_2} )  ) }^2{{\sigma }_3})Z^{[3,1]}
 \nonumber \\
 && \qquad - C_{[1,2]}( ( 3 + {{\sigma }_2} )( J^{[3,0]} - {\hat{\sigma}_1}
   ( 3{\sigma }_1^2 - {( 3 + {{\sigma }_2} ) }^2 )) +
  {\hat{\sigma}_1}{{\sigma }_1}( {\sigma }_1^2 - 4( 3 + {{\sigma }_2} )  ) {{\sigma
  }_3})Z^{[3,1]}\nonumber \\
  && \qquad+ B_{[2,0]}( {\hat{\sigma}_1} + {{\sigma }_1} )[ 3 K^{[3,0]}( 4{\hat{\sigma}_1}  + {{\sigma }_1} )  -
    J^{[3,0]}\hat{\sigma}_1^2( 3 + {{\sigma }_2} ) - 12{{\hat{\sigma}_1} }{{\sigma }_1}{{\sigma }_3}
    K^{[2,0]}  \nonumber \\&&\qquad
- {{\hat{\sigma}_1} }{{\sigma }_1}{{\sigma }_3}
     ( \sigma_1^2( 3 + {{\sigma }_2} )  -
       4{{\sigma }_1}{{\sigma }_3} )  ]+ B_{[0,1]}[9 K^{[3,0]}\hat{\sigma}_1^4 + 3{\sigma }_1^2( 3 - {\sigma }_1^2 )
   ( 3{\sigma }_1^2 - {( 3 + {{\sigma }_2} ) }^2 ) \nonumber \\
   && \qquad-
  3{\hat{\sigma}_1}{{\sigma }_1}( 18{\sigma }_1^4 + {( 3 + {{\sigma }_2} ) }^2( 15 + 4{{\sigma }_2} )  -
     3{\sigma }_1^2( 33 + 16{{\sigma }_2} + 2{\sigma }_2^2 )  )  \nonumber \\
     &&\qquad +
  \hat{\sigma}_1^2( 3{( 3 + {{\sigma }_2} ) }^3( -4 + 3{{\sigma }_2} )  +
     3{\sigma }_1^4( -21 + 8{{\sigma }_2} ) +
     {\sigma }_1^2( 297 + 9{{\sigma }_2} - 54{\sigma }_2^2 - 8{\sigma }_2^3 )  ) \nonumber \\
     &&\qquad  -
  3{{\sigma }_1}( {\sigma }_1^4 + {\sigma }_1^2( -9 + {{\sigma }_2} )  +
     4{( 3 + {{\sigma }_2} ) }^2 ) {{\sigma }_3} -
  2( {\sigma }_1^4 + 4{( 3 + {{\sigma }_2} ) }^2 - 2{\sigma }_1^2( 9 + 2{{\sigma }_2} )  )
   {\sigma }_3^2 \nonumber \\
     &&\qquad+ {\hat{\sigma}_1}{{\sigma }_3}( - {\sigma }_1^4{{\sigma }_2}  +
     {( 3 + {{\sigma }_2} ) }^2( 15 + {{\sigma }_2} )  - 18{\sigma }_1^2( 7 + 3{{\sigma }_2} )  +
    {\sigma }_1^3{{\sigma }_3}( 5{\sigma }_1^2 - 4{{\sigma }_2}))\nonumber \\
     &&\qquad-
  \hat{\sigma}_1^2{{\sigma }_3}( 2{\sigma }_1^5 + {\sigma }_1^3( 9 - 19{{\sigma }_2} )  +
     11{{\sigma }_1}{( 3 + {{\sigma }_2} ) }^2 + 8{\sigma }_1^2{{\sigma }_3} +
     4( 3 + {{\sigma }_2} ) {{\sigma }_3} )  \nonumber \\
     &&\qquad +
  \hat{\sigma}_1^3( {{\sigma }_1}( -33 + 9{\sigma }_1^2 - 10{{\sigma }_2} )
      ( 3{\sigma }_1^2 - {( 3 + {{\sigma }_2} ) }^2 )  \nonumber \\
     &&\qquad+
     {{\sigma }_3}( 7{\sigma }_1^4 - 12{( 3 + {{\sigma }_2} ) }^2 +
        3{\sigma }_1^2( 5 + 3{{\sigma }_2} )  - 4{{\sigma }_1}{{\sigma }_3} )
        )],
        \end{eqnarray}
        \begin{eqnarray}
&& Q_{4}^{\hat{\mathcal{O}}_2|2,2}(\{y\}_2,\{y'\}_2) =
9{\hat{\sigma}_1}( K^{[0,2]}{{\hat{\sigma}_1}}^2  -
K^{[0,2]}{\hat{\sigma}_2} - {{\hat{\sigma}_2}}^2 )
  9 K^{[0,2]}( 2{{\hat{\sigma}_1}}^2 - 3{\hat{\sigma}_2} ) {{\sigma }_1}\nonumber \\
  &&\qquad -3 {\hat{\sigma}_2}( -18 + 6{{\hat{\sigma}_1}}^2 + {{\hat{\sigma}_2}}^2   )
   {{\sigma }_1}
   + 3K^{[0,2]}{\hat{\sigma}_1}( K^{[0,2]} - {\hat{\sigma}_2} ) {\sigma }_1^2 \nonumber \\
   &&\qquad+
  ( -6K^{[0,2]}{{\hat{\sigma}_1}}^2 + {{\hat{\sigma}_2}}^2( 6 + {\hat{\sigma}_2} )  ) {\sigma }_1^3 +
  {\hat{\sigma}_1}( -3K^{[0,2]} + {\hat{\sigma}_2}( K^{[0,2]} + {\hat{\sigma}_2} )  ) {\sigma }_1^4 \nonumber \\
  &&\qquad+
  K^{[0,2]}{\hat{\sigma}_2}{\sigma }_1^5 +  {{\sigma }_1}
      ( 27 + 27{\hat{\sigma}_2} + 9{{\hat{\sigma}_2}}^2 + 4{{\hat{\sigma}_2}}^3 - 9{\sigma }_1^2 - 9{\hat{\sigma}_2}{\sigma }_1^2
      - \hat{\sigma}_2^2{\sigma }_1^2
        ) \nonumber \\
        &&\qquad + {\hat{\sigma}_1}( 9K^{[0,2]} + 3( 3\hat{\sigma}_1^2 + \hat{\sigma}_2^2 )  -
        ( 9 + 6{\hat{\sigma}_2} + 4\hat{\sigma}_2^2 ) {\sigma }_1^2 + ( K^{[0,2]} + {{\hat{\sigma}_1}}^2 ) {\sigma }_1^4 ) \nonumber \\
        &&\qquad +
      {{\hat{\sigma}_1}}^3( -( {\sigma }_1^2( -3 + {\sigma }_1^2 )  )  +
        3{\hat{\sigma}_2}( -1 + {\sigma }_1^2 )  )  -
     {{\hat{\sigma}_1}}^4( 3{\hat{\sigma}_1} + {{\sigma }_1}( 3 - {\hat{\sigma}_2} + {\sigma }_1^2 )  )  \nonumber \\
     &&\qquad +
     {{\hat{\sigma}_1}}^2{{\sigma }_1}( -5{{\hat{\sigma}_2}}^2 + 6{\sigma }_1^2 + {\hat{\sigma}_2}( -12 + 5{\sigma }_1^2 )  )
   {{\sigma }_2} + {\sigma }_2^2( -3{\hat{\sigma}_1}( 2 + {{\hat{\sigma}_1}}^2 - {\hat{\sigma}_2} ) {\hat{\sigma}_2}) \\
   &&\qquad +{\sigma }_2^2(-
     {\hat{\sigma}_2}{{\sigma }_1}( 6 + 5{\hat{\sigma}_2} - 3{\hat{\sigma}_1}{{\sigma }_1} )  +
     K^{[0,2]}( 3{\hat{\sigma}_1} + {\sigma }_1^3 )  - {\hat{\sigma}_2}( 3{\hat{\sigma}_1} - {{\sigma }_1} ) {{\sigma }_2} ) \nonumber\\ &&\qquad +
  K^{[0,2]}Z^{[2,2]}[ ( -1 + {\hat{A}_{[3,0]}} + {C_{[0,3]}} ) {{\sigma }_1}( -3 + {\sigma }_1^2 )  +
     ( {B_{[0,3]}} + {C_{[0,3]}}{\hat{\sigma}_1} ) {{\sigma }_2}]\nonumber \\ &&\qquad -Z^{[2,2]}[K^{[0,2]}K^{[2,0]}
     ( {\hat{A}_{[2,2]}} + B_{[2,2]}{\hat{\sigma}_1} )
      -
  {\hat{A}_{[3,0]}}( K^{[0,2]}{{\sigma }_1}( 3 - {\sigma }_1^2 )   +
     {\hat{\sigma}_2}( {\hat{\sigma}_1} + {{\sigma }_1} ) K^{[2,0]}
     )],\nonumber
\end{eqnarray}
and
\begin{eqnarray}
  && Q_{4}^{\hat{\mathcal{O}}_{2}|0,4}(\{y\}_3,y'_1)=( {\hat{A}_{[0,4]}} + {B_{[0,4]}}{{\hat{\sigma} }_1})K^{[0,4]} \nonumber
  \\&&\qquad -
  ( {\hat{A}_{[2,0]}} + {B_{[1,2]}}{{\hat{\sigma} }_1} ) {{\hat{\sigma} }_4}
   {( 27 - 6{{{\hat{\sigma} }_1}}^2 + 12{{\hat{\sigma} }_2} + {{{\hat{\sigma} }_2}}^2 - 2{{\hat{\sigma} }_1}{{\hat{\sigma} }_3}
   + 2{{\hat{\sigma} }_4} ) }^2 \nonumber \\&&\qquad +
  {B_{[0,1]}}[ 3{{\hat{\sigma} }_1}( 9{{{\hat{\sigma} }_1}}^4( 15 + {{\hat{\sigma} }_2} )  +
        2{( 3 + {{\hat{\sigma} }_2} ) }^3( 27 + {{\hat{\sigma} }_2}( 6 + {{\hat{\sigma} }_2} )  ) \nonumber \\&&\qquad
        -
        9{{{\hat{\sigma} }_1}}^2( 3 + {{\hat{\sigma} }_2} ) ( 33 + {{\hat{\sigma} }_2}( 10 + {{\hat{\sigma} }_2} )  )  )  +
      6( -4{( 3 + {{\hat{\sigma} }_2} ) }^4 + 3{{{\hat{\sigma} }_1}}^4( 21 + {{\hat{\sigma} }_2} )) {{\hat{\sigma} }_3}
      \nonumber \\&&\qquad - 12
        {\hat{\sigma} }_1^2( 54 + 27{{\hat{\sigma} }_2} + 6{\hat{\sigma} }_2^2 + {\hat{\sigma} }_2^3   ) {{\hat{\sigma} }_3} +
     {{\hat{\sigma} }_1}( 270 + 198{{\hat{\sigma} }_2} + 30{\hat{\sigma} }_2^2 - 2{\hat{\sigma} }_2^3 +
        3{\hat{\sigma} }_1^2( 39 + {{\hat{\sigma} }_2} )  ) {\hat{\sigma} }_3^2 \nonumber \\&&\qquad +
     4( 3{\hat{\sigma} }_1^2 + 2{( 3 + {{\hat{\sigma} }_2} ) }^2 ) {\hat{\sigma} }_3^3 +
     3{{\hat{\sigma} }_1}( 36{\hat{\sigma} }_1^4 - {\hat{\sigma} }_1^2
        ( 585 + {{\hat{\sigma} }_2}( 180 + 7{{\hat{\sigma} }_2} )  )){{\hat{\sigma} }_4}  \nonumber \\&&\qquad +
        3{{\hat{\sigma} }_1} ( 3 + {{\hat{\sigma} }_2} ) ( 567 + {{\hat{\sigma} }_2}
            ( 231 + {{\hat{\sigma} }_2}( 25 + {{\hat{\sigma} }_2} )  )  )   {{\hat{\sigma} }_4}+
     4{{\hat{\sigma} }_1}( -30 + 3{{{\hat{\sigma} }_1}}^2 - 7{{\hat{\sigma} }_2} ) {{{\hat{\sigma} }_3}}^2{{\hat{\sigma} }_4}\nonumber \\&&\qquad +
     3( 234 + 24{{{\hat{\sigma} }_1}}^4 + 114{{\hat{\sigma} }_2} + 6{{{\hat{\sigma} }_2}}^2 - 2{{{\hat{\sigma} }_2}}^3 -
        {{{\hat{\sigma} }_1}}^2( 303 + 80{{\hat{\sigma} }_2} + {{{\hat{\sigma} }_2}}^2 )  ) {{\hat{\sigma} }_3}{{\hat{\sigma} }_4} \nonumber \\&&\qquad +
     ( -4{{{\hat{\sigma} }_1}}^3( 12 + {{\hat{\sigma} }_2} )  +
        3{{\hat{\sigma} }_1}( 168 + 63{{\hat{\sigma} }_2} + 5{{{\hat{\sigma} }_2}}^2 )  - 12{{{\hat{\sigma} }_1}}^2{{\hat{\sigma} }_3} +
        24( 4 + {{\hat{\sigma} }_2} ) {{\hat{\sigma} }_3} ) {{{\hat{\sigma} }_4}}^2 ]  \nonumber \\&&\qquad +
  {\hat{A}_{[1,2]}}  [3( -9{{{\hat{\sigma} }_1}}^6 - 18{{{\hat{\sigma} }_1}}^2{( 3 + {{\hat{\sigma} }_2} ) }^2
         ( 5 + {{\hat{\sigma} }_2} )  + {( 3 + {{\hat{\sigma} }_2} ) }^4( 15 + {{\hat{\sigma} }_2} )  +
        3{{{\hat{\sigma} }_1}}^4( 54 + 21{{\hat{\sigma} }_2} + {{{\hat{\sigma} }_2}}^2 )  ) \nonumber\\&&\qquad -
     3{{\hat{\sigma} }_1}( 9{{{\hat{\sigma} }_1}}^4 + 6{( 3 + {{\hat{\sigma} }_2} ) }^2( 7 + {{\hat{\sigma} }_2} )  -
        {{{\hat{\sigma} }_1}}^2( 135 + 48{{\hat{\sigma} }_2} + {{{\hat{\sigma} }_2}}^2 )  ) {{\hat{\sigma} }_3}\nonumber\\&&\qquad -
     ( 9{{{\hat{\sigma} }_1}}^4 - 39{{{\hat{\sigma} }_1}}^2( 3 + {{\hat{\sigma} }_2} )  +
        {( 3 + {{\hat{\sigma} }_2} ) }^2( 15 + {{\hat{\sigma} }_2} )  ) {{{\hat{\sigma} }_3}}^2 +
     {{\hat{\sigma} }_1}( -{{{\hat{\sigma} }_1}}^2 + 4( 3 + {{\hat{\sigma} }_2} )  ) {{{\hat{\sigma} }_3}}^3 \nonumber\\&&\qquad+
     ( 3{{{\hat{\sigma} }_1}}^4( -3 + 4{{\hat{\sigma} }_2} )  +
        2{( 3 + {{\hat{\sigma} }_2} ) }^2( 36 + 21{{\hat{\sigma} }_2} + {{{\hat{\sigma} }_2}}^2 )  -
        {{{\hat{\sigma} }_1}}^2( 135 + 153{{\hat{\sigma} }_2} + 39{{{\hat{\sigma} }_2}}^2 + {{{\hat{\sigma} }_2}}^3 )  ) {{\hat{\sigma} }_4}\nonumber\\&&\qquad +
     {{\hat{\sigma} }_1}( -27 - 24{{\hat{\sigma} }_2} - 5{{{\hat{\sigma} }_2}}^2 + 2{{{\hat{\sigma} }_1}}^2( -9 + 2{{\hat{\sigma} }_2} )  )
      {{\hat{\sigma} }_3}{{\hat{\sigma} }_4} - ( 5{{{\hat{\sigma} }_1}}^2 + 4( 3 + {{\hat{\sigma} }_2} )  ) {{{\hat{\sigma} }_3}}^2
      {{\hat{\sigma} }_4}\nonumber\\&&\qquad+ ( -54 + {{{\hat{\sigma} }_1}}^2( 27 - 4{{\hat{\sigma} }_2} )  - 3{{\hat{\sigma} }_2}
      + 5{{{\hat{\sigma} }_2}}^2 +
        13{{\hat{\sigma} }_1}{{\hat{\sigma} }_3} - 7{{\hat{\sigma} }_4} ) {{{\hat{\sigma} }_4}}^2 ] \nonumber\\&&\qquad +
  ( {B_{[2,0]}} - {B_{[1,2]}} ) [ -3{{\hat{\sigma} }_1}
      ( 108{{{\hat{\sigma} }_1}}^4 - 3{{{\hat{\sigma} }_1}}^2( 3 + {{\hat{\sigma} }_2} ) {( 9 + {{\hat{\sigma} }_2} ) }^2 +
        {( 3 + {{\hat{\sigma} }_2} ) }^3( 45 + {{\hat{\sigma} }_2}( 6 + {{\hat{\sigma} }_2} )  )  ) \nonumber\\&&\qquad +
     6( -54{{{\hat{\sigma} }_1}}^4 + 2{( 3 + {{\hat{\sigma} }_2} ) }^4 +
        {{{\hat{\sigma} }_1}}^2( 3 + {{\hat{\sigma} }_2} ) ( 45 + {{\hat{\sigma} }_2}( 6 + {{\hat{\sigma} }_2} )  )  )
      {{\hat{\sigma} }_3}+
     4{{\hat{\sigma} }_1}( -2{{{\hat{\sigma} }_1}}^2 + 3( 5 + {{\hat{\sigma} }_2} )  ) {{{\hat{\sigma} }_3}}^2{{\hat{\sigma} }_4} \nonumber\\&&\qquad
     + {{\hat{\sigma} }_1}( -108{{{\hat{\sigma} }_1}}^2 +
        ( 3 + {{\hat{\sigma} }_2} ) ( -27 + ( -18 + {{\hat{\sigma} }_2} ) {{\hat{\sigma} }_2} )  ) {{{\hat{\sigma} }_3}}^2 -
     4( 3{{{\hat{\sigma} }_1}}^2 + {( 3 + {{\hat{\sigma} }_2} ) }^2 ) {{{\hat{\sigma} }_3}}^3\nonumber\\&&\qquad -
     {{\hat{\sigma} }_1}( 3483 + 72{{{\hat{\sigma} }_1}}^4 + 2565{{\hat{\sigma} }_2} + 603{{{\hat{\sigma} }_2}}^2 + 51{{{\hat{\sigma} }_2}}^3 +
        2{{{\hat{\sigma} }_2}}^4 - 12{{{\hat{\sigma} }_1}}^2( 99 + 30{{\hat{\sigma} }_2} + {{{\hat{\sigma} }_2}}^2 )  ) {{\hat{\sigma} }_4}\nonumber\\&&\qquad +
     ( -48{{{\hat{\sigma} }_1}}^4 + 4{{{\hat{\sigma} }_1}}^2( 144 + 39{{\hat{\sigma} }_2} + {{{\hat{\sigma} }_2}}^2 )  +
        3( -117 - 57{{\hat{\sigma} }_2} - 3{{{\hat{\sigma} }_2}}^2 + {{{\hat{\sigma} }_2}}^3 )  ) {{\hat{\sigma} }_3}{{\hat{\sigma} }_4} \nonumber\\&&\qquad-
     4{{{\hat{\sigma} }_4}}^2( -12{{{\hat{\sigma} }_1}}^3 - 4{{{\hat{\sigma} }_1}}^2{{\hat{\sigma} }_3} +
        3( 4 + {{\hat{\sigma} }_2} ) {{\hat{\sigma} }_3} +
        {{\hat{\sigma} }_1}( 99 + 2{{\hat{\sigma} }_2}( 18 + {{\hat{\sigma} }_2} )  + 2{{\hat{\sigma} }_4} )  )  ] \nonumber\\&&\qquad +
  {C_{[1,2]}}[ -9{{\hat{\sigma} }_1}( 3{{{\hat{\sigma} }_1}}^6 - 4{{\hat{\sigma} }_2}{( 3 + {{\hat{\sigma} }_2} ) }^3 -
        {{{\hat{\sigma} }_1}}^4( 18 + 21{{\hat{\sigma} }_2} + {{{\hat{\sigma} }_2}}^2 )  +
        {{{\hat{\sigma} }_1}}^2( 27 + 99{{\hat{\sigma} }_2} + 45{{{\hat{\sigma} }_2}}^2 + 5{{{\hat{\sigma} }_2}}^3 )  )\nonumber \\&&\qquad -
     3( 9{{{\hat{\sigma} }_1}}^6 - 4{( 3 + {{\hat{\sigma} }_2} ) }^4 -
        {{{\hat{\sigma} }_1}}^4( 27 + 48{{\hat{\sigma} }_2} + {{{\hat{\sigma} }_2}}^2 )  +
        4{{{\hat{\sigma} }_1}}^2( 27 + 45{{\hat{\sigma} }_2} + 15{{{\hat{\sigma} }_2}}^2 + {{{\hat{\sigma} }_2}}^3 )  ) {{\hat{\sigma} }_3}\nonumber\\&&\qquad -
     3{{\hat{\sigma} }_1}( 3{{{\hat{\sigma} }_1}}^4 - {{{\hat{\sigma} }_1}}^2( 3 + 13{{\hat{\sigma} }_2} )  +
        12( 6 + 5{{\hat{\sigma} }_2} + {{{\hat{\sigma} }_2}}^2 )  ) {{{\hat{\sigma} }_3}}^2 -
     ( {{{\hat{\sigma} }_1}}^4 - 4{{{\hat{\sigma} }_1}}^2{{\hat{\sigma} }_2} + 4{( 3 + {{\hat{\sigma} }_2} ) }^2 ) {{{\hat{\sigma} }_3}}^3 \nonumber\\&&\qquad
     +
{{\hat{\sigma} }_1}( 3{{{\hat{\sigma} }_1}}^4( -15 +
4{{\hat{\sigma} }_2} )  -
        {{{\hat{\sigma} }_1}}^2( -729 - 63{{\hat{\sigma} }_2} + 39{{{\hat{\sigma} }_2}}^2 + {{{\hat{\sigma} }_2}}^3 )) {{\hat{\sigma} }_4}\nonumber\\&&\qquad +
        {{\hat{\sigma} }_1}( 3 + {{\hat{\sigma} }_2} ) ( -702 + {{\hat{\sigma} }_2}
            ( -135 + {{\hat{\sigma} }_2}( 24 + {{\hat{\sigma} }_2} )  )  )   {{\hat{\sigma} }_4} +
     {{\hat{\sigma} }_1}( -9{{{\hat{\sigma} }_1}}^2 + 8( 6 + {{\hat{\sigma} }_2} )  ) {{{\hat{\sigma} }_3}}^2{{\hat{\sigma} }_4} \nonumber\\&&\qquad+
     ( {{{\hat{\sigma} }_1}}^4( -42 + 4{{\hat{\sigma} }_2} )  +
        {{{\hat{\sigma} }_1}}^2( 441 + 84{{\hat{\sigma} }_2} - 5{{{\hat{\sigma} }_2}}^2 )  +
        3( -117 - 57{{\hat{\sigma} }_2} - 3{{{\hat{\sigma} }_2}}^2 + {{{\hat{\sigma} }_2}}^3 )  ) {{\hat{\sigma} }_3}{{\hat{\sigma} }_4}\nonumber \\&&\qquad-
     {{{\hat{\sigma} }_4}}^2( {{{\hat{\sigma} }_1}}^3( -51 + 4{{\hat{\sigma} }_2} )  - 21{{{\hat{\sigma} }_1}}^2{{\hat{\sigma} }_3} +
        12( 4 + {{\hat{\sigma} }_2} ) {{\hat{\sigma} }_3} +
        {{\hat{\sigma} }_1}( 342 + 99{{\hat{\sigma} }_2} - {{{\hat{\sigma} }_2}}^2 + 11{{\hat{\sigma} }_4} )  )
        ].
\end{eqnarray}

\subsubsection{From boundary to bulk form factors}
The bulk counterparts of the solutions above are:
\begin{eqnarray}
   Q_{2;\text{bulk}}^{\hat{\mathcal{O}}_{2}|1,1}(y_1,y'_1)&=& \hat{\sigma}_1 (\sigma_1 +\hat{\sigma}_1)  , \label{q11}\\
    Q_{2;\text{bulk}}^{\hat{\mathcal{O}}_{2}|2,0}(y_1,y_2)&=&   \sigma_1 (B_{[0,1]}\sigma_1^2+ B_{[2,0]}
   K^{[2,0]}_{\text{bulk}}),\\
 Q_{3;\text{bulk}}^{\hat{\mathcal{O}}_{2}|1,2}(y_1,y'_1,y'_2) &=& [(B_{[1,2]}-B_{[2,0]})\sigma_1+C_{[1,2]}\hat{\sigma}_1]Z^{[1,2]}K_{\text{bulk}}^{[0,2]}
  -\hat{\sigma}_2 B_{[2,0]}(\sigma_1+\hat{\sigma}_1) Z^{[1,2]}\nonumber \\
  &+& B_{[0,1]}[(\sigma_1+\hat{\sigma}_1)(\hat{\sigma}_2^2-2\sigma_1^2 \hat{\sigma}_1^2 + 3 \hat{\sigma}_2 \sigma_1^2)+ 2 \hat{\sigma}_2 \sigma_1
  (\hat{\sigma}_2 + \sigma_1 \hat{\sigma}_1)], \label{q122}\\
Q_{3;\text{bulk}}^{\hat{\mathcal{O}}_{2}|3,0}(\{y\}_3)&=&\sigma_2^2(\sigma_1
\sigma_2-4\sigma_3)+ \sigma_1\hat{A}_{[3,0]}
K_{\text{bulk}}^{[3,0]}, \\
Q_{3;\text{bulk}}^{\hat{\mathcal{O}}_{2}|0,3}(\{y'\}_3)&=&4\hat{\sigma}_2^2\hat{\sigma}_3+
\hat{\sigma}_1 C_{[0,3]}
K_{\text{bulk}}^{[0,3]},\\
Q_{4;\text{bulk}}^{\hat{\mathcal{O}}_{2}|3,1}(\{y\}_3, y'_1)&=&(
B_{[3,1]}\sigma_1
+C_{[3,1]}\hat{\sigma}_1)K^{[3,0]}_{\text{bulk}}Z^{[3,1]}
\nonumber \\
&+&  ( B_{[2,0]} - {B_{[1,2]}} ) ( {( {\sigma }_1^2 - 2\,{{\sigma
}_2} ) }^2{{\sigma }_3} -
     4 K^{[3,0]}_{\text{bulk}}{{\hat{\sigma} }_1} ) Z^{[3,1]} \nonumber \\
&-&  {C_{[1,2]}}
   ( {{\sigma }_2^2}(\sigma_1 \sigma_2 - 4 \sigma_3) + {\sigma }_2^3{{\hat{\sigma} }_1} +
     {{\sigma }_1}( {\sigma }_1^2 - 4{{\sigma }_2} ) {{\sigma }_3}{{\hat{\sigma} }_1} )Z^{[3,1]}  \nonumber \\
&-&
  {B_{[2,0]}}{{\hat{\sigma} }_1} ( {{\sigma }_1} + {{\hat{\sigma}}_1} )( {{\sigma }_1}{{\sigma }_2} - 4{{\sigma }_3} ) (  {\sigma }_1^2
         {{\sigma }_3}  + {{\sigma }_2^2}{\hat{\sigma}}_1
     )   \nonumber \\
&+& {B_{[0,1]}}[ -2{( {\sigma }_1^2 - 2{{\sigma }_2} ) }^2{\sigma
}_3^2 +
     {{\sigma }_3}( - {\sigma }_1^4{{\sigma }_2}  + {\sigma }_2^3 +
        {{\sigma }_1}( 5{\sigma }_1^2 - 4{{\sigma }_2} ) {{\sigma }_3} ) {{\hat{\sigma} }_1}  \\
        & + &
     ( {{\sigma }_1}{\sigma }_2^3( 9{{\sigma }_2} -8{\sigma }_1^2  )
         (19{\sigma }_1^2{{\sigma }_2} - 11{\sigma }_2^2 -2{\sigma }_1^4) {{\sigma }_3} -
        4( 2{\sigma }_1^2 + {{\sigma }_2} ) {\sigma }_3^2 ) {\hat{\sigma}}_1^2\nonumber \\
        & + &
     ( {{\sigma }_1}{\sigma }_2^2( 10{{\sigma }_2}-9{\sigma }_1^2  )  +
        {{\sigma }_3}( 7{\sigma }_1^4 + 9{\sigma }_1^2{{\sigma }_2} - 12{\sigma }_2^2 -
           4{{\sigma }_1}{{\sigma }_3} )  +9 K^{[3,0]}_{\text{bulk}}{\hat{\sigma} }_1 ) {\hat{\sigma}}_1^3
           ],\nonumber\\
Q_{4;\text{bulk}}^{\hat{\mathcal{O}}_{2}|0,4}(\{y'\}_4)&=&{B_{[0,4]}}{{\hat{\sigma}
}_1}K^{[0,4]}_{\text{bulk}}- {B_{[1,2]}}{{\hat{\sigma} }_1}
{{\hat{\sigma} }_4}
   {( {{{\hat{\sigma} }_2}}^2 - 2{{\hat{\sigma} }_1}{{\hat{\sigma} }_3}
   + 2{{\hat{\sigma} }_4} ) }^2 \nonumber \\&+ &
   (B_{[2,0]}-B_{[1,2]})[{\hat{\sigma} }_2^2( {{\hat{\sigma} }_1}{{\hat{\sigma} }_2} - 4{{\hat{\sigma} }_3} ) {\hat{\sigma} }_3^2 -
  2{{\hat{\sigma} }_1}{\hat{\sigma} }_2^4{{\hat{\sigma} }_4} + {\hat{\sigma} }_2^2( 4{\hat{\sigma} }_1^2 + 3{{\hat{\sigma} }_2} ) {{\hat{\sigma} }_3}
   {{\hat{\sigma} }_4}\nonumber \\
   & - & 4{{\hat{\sigma} }_1}( 2{\hat{\sigma} }_1^2 - 3{{\hat{\sigma} }_2} ) {\hat{\sigma} }_3^2{{\hat{\sigma} }_4} +
  4( 4{\hat{\sigma} }_1^2 - 3{{\hat{\sigma} }_2} ) {{\hat{\sigma} }_3}{\hat{\sigma} }_4^2
  -
  8{{\hat{\sigma} }_1}({\hat{\sigma} }_2^2 +{\hat{\sigma}
  }_4){\hat{\sigma} }_4^2]\nonumber \\
  & + & B_{[0,1]}[-2{\hat{\sigma} }_2^2( {{\hat{\sigma} }_1}{{\hat{\sigma} }_2} - 4{{\hat{\sigma} }_3} ) {\hat{\sigma} }_3^2 +
  ( 3{{\hat{\sigma} }_1}{\hat{\sigma} }_2^4 - 3{\hat{\sigma} }_2^2( {\hat{\sigma} }_1^2 + 2{{\hat{\sigma} }_2} ) {{\hat{\sigma} }_3}
  \\&+&
     4{{\hat{\sigma} }_1}( 3{{{\hat{\sigma} }_1}}^2 - 7{{\hat{\sigma} }_2} ) {\hat{\sigma} }_3^2 ) {{\hat{\sigma} }_4} -
  ( {{\hat{\sigma} }_1}( 4{\hat{\sigma} }_1^2 - 15{{\hat{\sigma} }_2} ) {{\hat{\sigma} }_2} +
     12( {\hat{\sigma} }_1^2 - 2{{\hat{\sigma} }_2} ) {{\hat{\sigma} }_3} ) {\hat{\sigma}
     }_4^2]\nonumber \\
     &+ & C_{[1,2]}[- {( {\hat{\sigma} }_1^2 - 2{{\hat{\sigma} }_2} ) }^2{\hat{\sigma} }_3^3  -
  {{\hat{\sigma} }_1}( {\hat{\sigma} }_1^2 - {{\hat{\sigma} }_2} ) {\hat{\sigma} }_2^3{{\hat{\sigma} }_4} +
  {{\hat{\sigma} }_3}( {{\hat{\sigma} }_2}( 4{\hat{\sigma} }_1^4 - 5{\hat{\sigma} }_1^2{{\hat{\sigma} }_2} + 3{\hat{\sigma} }_2^2 )\nonumber \\
  &  - &
     {{\hat{\sigma} }_1}( 9{\hat{\sigma} }_1^2 - 8{{\hat{\sigma} }_2} ) {{\hat{\sigma} }_3} ) {{\hat{\sigma} }_4} -
  {\hat{\sigma} }_4^2( {{\hat{\sigma} }_1}( 4{\hat{\sigma} }_1^2 - {{\hat{\sigma} }_2} ) {{\hat{\sigma} }_2} -
     3( 7{\hat{\sigma} }_1^2 - 4{{\hat{\sigma} }_2} ) {{\hat{\sigma} }_3} + 11{{\hat{\sigma} }_1}{{\hat{\sigma} }_4} )
     ],\nonumber \\
   Q_{4;\text{bulk}}^{\hat{\mathcal{O}}_{2}|2,2}(\{y\}_2,\{y'\}_2)&=&{\hat{\sigma}_2}{\sigma }_1^3( K^{[0,2]}{\sigma }_1^2 +
     \hat{\sigma}_2( {\hat{\sigma}_2} + {\hat{\sigma}_1}{{\sigma }_1} ))
     +
  4{{\hat{\sigma}_2}}^2{{\sigma }_1}( {\hat{\sigma}_2} - {\hat{\sigma}_1}{{\sigma }_1} ) {{\sigma }_2} \nonumber \\
  &+&
  {\hat{\sigma}_2}{{\sigma }_1}( {{\hat{\sigma}_1}}^4 - {\hat{\sigma}_2}{\sigma }_1^2 ) {{\sigma }_2}
  + 3{\hat{\sigma}_1}{{\hat{\sigma}_2}}^2{\sigma }_2^2 +
  3{\hat{\sigma}_1}{\hat{\sigma}_2}( {\sigma }_1^2 - {{\sigma }_2} ) {{\sigma }_2}
  ( {{\hat{\sigma}_1}}^2 + {{\sigma }_2} ) \nonumber \\ &+&
 K^{[0,2]}{\hat{\sigma}_1}{\sigma }_1^4( {\hat{\sigma}_2} + {{\sigma }_2} )
 -
  5{\hat{\sigma}_2}{{\sigma }_1}{{\sigma }_2}(  {{\hat{\sigma}_1}}^2({\hat{\sigma}_2} -{\sigma
  }_1^2)
  + {\hat{\sigma}_2}{{\sigma }_2} ) \nonumber \\ &-&
   {{\sigma }_1}{{\sigma }_2}( {{\hat{\sigma}_1}}^4{\sigma }_1^2 - {\hat{\sigma}_2}{\sigma }_2^2 ) +
 Z^{[2,2]}{\hat{A}_{[3,0]}}(
     K^{[2,0]}{\hat{\sigma}_2}( {\hat{\sigma}_1} + {{\sigma }_1} ) - K^{[0,2]}{\sigma }_1^3   )\nonumber \\&
      +&
 K^{[0,2]}Z^{[2,2]}( ({\hat{A}_{[3,0]}} + {C_{[0,3]}} -1) {\sigma }_1^3 +
     {\hat{\sigma}_1}({C_{[0,3]}}{{\sigma }_2} - K^{[2,0]}{{B}_{[2,2]}} )
     ).\label{news}
\end{eqnarray}
Equations (\ref{q11})-(\ref{news}) provide a set of solutions to
the bulk form factor equations hitherto unknown.

\section{Conclusions and outlook}\label{s4}

\noindent In this paper we have initiated the boundary form factor
program for theories with many particles, concentrating on the
example of $A_n$-ATFTs. We have computed all minimal one- and
two-particle form factors and provided a description of the pole
structure of higher particle form factors for these theories. We
have then specialize our study to the self-dual point, $B=1$, and
to the $A_2$-theory, which possesses a pair of particles,
$1=\bar{2}$, that can be regarded as bound states resulting from
the processes $1+1 \rightarrow 2$ and $2+2 \rightarrow 1$. The
presence of bound states implies that besides the kinematic
residue equations also the bound state residue equations need to
be satisfied, which makes the computation of form factors a lot
more involved.

For the $A_2$-case we have obtained all form factors, up to four
particles, of two families of fields which correspond to spinless
and spin-1 fields of the bulk theory, respectively. Indeed we have
shown that the form factors of our first family of solutions
reduce to those obtained in \cite{oota} for bulk spinless fields
in the appropriate limit. In addition, we have obtained all form
factors up to four particles of another class of fields which
correspond to spin-1 bulk fields other than simple derivatives of
the previous ones. The bulk form factors of these fields were not
known up to now and have been obtained as a certain limit of our
boundary solutions.

This work provides further support to the statement that the
boundary form factor program \cite{BPT} is a very useful tool for
the computation of form factors of boundary fields, even for
multi-particle theories. The fact remains however that the
structure of these solutions is a lot more involved than for the
bulk case and identifying general patterns is still a challenge,
even for single particle models.

We would like to finish by emphasizing that the boundary form
factor program for IQFTs is still at its early stages of
development. The study of more models should provide further
insight into the mathematical structures of the form factors and
applications of the program to the computation of correlation
functions should be further explored. Focusing on the present
work, a natural follow up would be to look at ATFTs related to
other simple Lie algebras, including the completion of the study
of the $A_n$-case which we hope to carry out in a near future.

\paragraph{Acknowledgments:}

The author is grateful to Benjamin Doyon and Andreas Fring for
their insightful comments and careful reading of this manuscript.

\small

\begin{thebibliography}{10}

\bibitem{Weisz}
P.~Weisz,
\newblock Exact quantum sine-Gordon soliton form-factors,
\newblock Phys. Lett. {\bf B67}, 179 (1977).

\bibitem{KW}
M.~Karowski and P.~Weisz,
\newblock Exact S matrices and form-factors in (1+1)-dimensional field
  theoretic models with soliton behavior,
\newblock Nucl. Phys. {\bf B139}, 455--476 (1978).

\bibitem{smirnovbook}
F.~Smirnov,
\newblock Form factors in completely integrable models of quantum field theory,
\newblock Adv. Series in Math. Phys. {\bf 14}, World Scientific, Singapore
  (1992).

\bibitem{Z}
A.~B. Zamolodchikov,
\newblock Two point correlation function in scaling Lee-Yang model,
\newblock Nucl. Phys. {\bf B348}, 619--641 (1991).

\bibitem{FMS}
A.~Fring, G.~Mussardo, and P.~Simonetti,
\newblock Form-factors for integrable Lagrangian field theories, the
  sinh-Gordon theory,
\newblock Nucl. Phys. {\bf B393}, 413--441 (1993).

\bibitem{Essler:2004ht}
F.~H.~L. E{\ss}ler and R.~M. Konik,
\newblock Applications of massive integrable quantum field theories to problems
  in condensed matter physics,
\newblock I.~Kogan Memorial Volume, Wold Scientific, cond-mat/0412421 .

\bibitem{Cherednik:1985vs}
I.~V. Cherednik,
\newblock Factorizing particles on a half line and root systems,
\newblock Theor. Math. Phys. {\bf 61}, 977--983 (1984).

\bibitem{Sklyanin:1988yz}
E.~K. Sklyanin,
\newblock Boundary conditions for integrable quantum systems,
\newblock J. Phys. {\bf A21}, 2375--2389 (1988).

\bibitem{Fring:1993wt}
A.~Fring and R.~K{\"o}berle,
\newblock Affine Toda field theory in the presence of reflecting boundaries,
\newblock Nucl. Phys. {\bf B419}, 647--664 (1994).

\bibitem{Fring:1994ci}
A.~Fring and R.~K{\"o}berle,
\newblock Boundary bound states in affine Toda field theory,
\newblock Int. J. Mod. Phys. {\bf A10}, 739--752 (1995).

\bibitem{Fring}
A.~Fring and R.~K{\"o}berle,
\newblock Boundary bound states in affine Toda field theory,
\newblock Int. J. Mod. Phys. {\bf A10}, 739--752 (1995).

\bibitem{Ghoshal:1993tm}
S.~Ghoshal and A.~B. Zamolodchikov,
\newblock Boundary S matrix and boundary state in two-dimensional integrable
  quantum field theory,
\newblock Int. J. Mod. Phys. {\bf A9}, 3841--3886 (1994).

\bibitem{Konik:1995ws}
R.~Konik, A.~LeClair, and G.~Mussardo,
\newblock On Ising correlation functions with boundary magnetic field,
\newblock Int. J. Mod. Phys. {\bf A11}, 2765--2782 (1996).

\bibitem{dirk}
D.~Schuricht and H.~L. E{\ss}ler,
\newblock Dynamical response functions in the quantum Ising chain with a
  boundary, arXiv:0709.1809.

\bibitem{BPT}
Z.~Bajnok, L.~Palla, and G.~Tak\'acs,
\newblock On the boundary form factor program,
\newblock Nucl. Phys. {\bf B750}, 179--212 (2006).

\bibitem{Jimbo:1994gm}
M.~Jimbo, R.~Kedem, H.~Konno, T.~Miwa, and R.~Weston,
\newblock Difference equations in spin chains with a boundary,
\newblock Nucl. Phys. {\bf B448}, 429--456 (1995).

\bibitem{Jimbo:1994ja}
M.~Jimbo, R.~Kedem, T.~Kojima, H.~Konno, and T.~Miwa,
\newblock XXZ chain with a boundary,
\newblock Nucl. Phys. {\bf B441}, 437--470 (1995).

\bibitem{Miwa:1996ht}
T.~Miwa and R.~Weston,
\newblock Boundary ABF models,
\newblock Nucl. Phys. {\bf B486}, 517--545 (1997).

\bibitem{Furutsu:1999ph}
H.~Furutsu and T.~Kojima,
\newblock $U_q(\hat{sl}_n)$-analog of the XXZ chain with a boundary,
\newblock J. Math. Phys. {\bf 41}, 4413--4436 (2000).

\bibitem{Furutsu:1999by}
H.~Furutsu, T.~Kojima, and Y.~H. Quano,
\newblock Form-factors of the SU(2) invariant massive Thirring model with
  boundary reflection,
\newblock Int. J. Mod. Phys. {\bf A15}, 3037--3052 (2000).

\bibitem{JMi}
M.~Jimbo and T.~Miwa,
\newblock Algebraic Analysis of Solvable Lattice Models,
\newblock American Math. Soc., Providence, RI  (1995).

\bibitem{JMi2}
M.~Jimbo and T.~Miwa,
\newblock Quantum KZ equation with $|Q|=1$ and correlation functions of the XXZ
  model in the gapless regime,
\newblock J. Phys. {\bf A29}, 2923--2958 (1996).

\bibitem{JMi3}
M.~Jimbo, K.~Miki, T.~Miwa, and A.~Nakayashiki,
\newblock Correlation functions of the XXZ model for $\Delta <-1$,
\newblock Phys. Lett. {\bf A168}, 256--263 (1992).

\bibitem{mypaper}
O.~A. Castro-Alvaredo,
\newblock Boundary form factors of the sinh-Gordon model with Dirichlet
  boundary conditions at the self-dual point,
\newblock J. Phys. {\bf A39}, 11901--11914 (2006).

\bibitem{st}
M.~Sz\~ots and G.~Tak\'acs,
\newblock Spectrum of local boundary operators from boundary form factor
  bootstrap,
\newblock hep-th/0703226 .

\bibitem{toda1}
A.~Mikhailov, M.~Olshanetsky, and A.~Perelomov,
\newblock Two-dimensional generalized Toda lattice,
\newblock Comm. Math. Phys. {\bf 79}, 473--488 (1981).

\bibitem{toda2}
D.~I. Olive and N.~Turok,
\newblock Local conserved densities and zero curvature conditions for Toda
  lattice field theories,
\newblock Nucl. Phys. {\bf B257}, 277--301 (1985).

\bibitem{P2CFT}
A.~B. Zamolodchikov,
\newblock Integrable field theory from conformal field theory,
\newblock Adv. Stud. Pure Math. {\bf 19}, 641--674 (1989).

\bibitem{PCFT}
A.~B. Zamolodchikov,
\newblock Integrals of motion and S Matrix of the (Scaled) T=T(c) Ising model
  with magnetic field,
\newblock Int. J. Mod. Phys. {\bf A4}, 4235 (1989).

\bibitem{BCDS}
H.~W. Braden, E.~Corrigan, P.~E. Dorey, and R.~Sasaki,
\newblock Affine Toda field theory and exact S-matrices,
\newblock Nucl. Phys. {\bf B338}, 689--746 (1990).

\bibitem{FKS}
A.~Fring, C.~Korff, and J.~Schulz,
\newblock On the universal representation of the scattering matrix of affine
  Toda field theory,
\newblock Nucl. Phys. {\bf B567 [FS]}, 409--453 (2000).

\bibitem{oota}
T.~Oota,
\newblock Functional equations of form factors for diagonal scattering
  theories,
\newblock Nucl. Phys. {\bf B466}, 361--382 (1996).

\bibitem{Babujian:2003za}
H.~Babujian and M.~Karowski,
\newblock Exact form factors for the scaling $\mathbb{Z}_n$-Ising and the
  affine $A_{N-1}$ Toda quantum field theories,
\newblock Phys. Lett. {\bf B575}, 144--150 (2003).

\bibitem{asmatrices}
A.~E. Arinshtein, V.~A. Fateev, and A.~B. Zamolodchikov,
\newblock Quantum S-matrix of the (1+1)-dimensional Toda chain,
\newblock Phys. Lett. {\bf B87}, 389--392 (1979).

\bibitem{Koberle:1979sg}
R.~K{\"o}berle and J.~A. Swieca,
\newblock Factorizable $\mathbb{Z}_N$ models,
\newblock Phys. Lett. {\bf B86}, 209--210 (1979).

\bibitem{corrigan}
E.~Corrigan, P.~E. Dorey, R.~H. Rietdijk, and R.~Sasaki,
\newblock Affine Toda field theory on a half line,
\newblock Phys. Lett. {\bf B333}, 83--91 (1994).

\bibitem{corrigan3}
P.~Bowcock, E.~Corrigan, P.~E. Dorey, and R.~H. Rietdijk,
\newblock Classically integrable boundary conditions for affine Toda field
  theories,
\newblock Nucl. Phys. {\bf B445}, 469--500 (1995).

\bibitem{corrigan2}
E.~Corrigan, P.~E. Dorey, and R.~H. Rietdijk,
\newblock Aspects of affine Toda field theory on a half line,
\newblock Prog. Theor. Phys. Suppl. {\bf 118}, 143--164 (1995).

\bibitem{reflections}
O.~Castro-Alvaredo and A.~Fring,
\newblock Universal boundary reflection amplitudes,
\newblock Nucl. Phys. {\bf B682}, 551--584 (2004).

\bibitem{garden2}
G.~M. Gardenberger and G.~W. Delius,
\newblock Particle reflection amplitudes in $a_{(1)}^{(n)}$ Toda field
  theories,
\newblock Nucl. Phys. {\bf B554}, 325--364 (1999).

\bibitem{garden1}
G.~M. Gardenberger,
\newblock On $a_{(1)}^{(2)}$ reflection matrices and affine Toda theories,
\newblock Nucl. Phys. {\bf B542}, 659--693 (1999).

\bibitem{fateev}
V.~A. Fateev,
\newblock Normalization factors, reflection amplitudes and integrable systems,
  hep-th/0103014.

\bibitem{sinh1}
S.~Ghoshal,
\newblock Bound state boundary $S$-matrix of the sinh-Gordon model,
\newblock Int. J. Mod. Phys. {\bf A9}, 4801--4810 (1994).

\bibitem{sinh2}
E.~Corrigan,
\newblock On duality and reflection factors for the sinh-Gordon model with a
  boundary,
\newblock Int. J. Mod. Phys. {\bf A13}, 2709--2722 (1998).

\bibitem{sinh3}
H.~S. Cho, K.~S. Soh, and J.~D. Kim,
\newblock One-loop boundary reflection for the integrable boundary sinh-Gordon
  model,
\newblock J. Korean Phys. Soc. {\bf 32}, 661--665 (1998).

\bibitem{sinh4}
E.~Corrigan and G.~W. Delius,
\newblock Boundary breathers in the sinh-Gordon model,
\newblock J. Phys. {\bf A32}, 8601--8614 (1999).

\bibitem{sinh5}
A.~Chenaghlou and E.~Corrigan,
\newblock First order quantum corrections to the classical reflection factor of
  the sinh-Gordon model,
\newblock Int. J. Mod. Phys. {\bf A15}, 4417--4432 (2000).

\bibitem{sinh6}
E.~Corrigan and A.~Taormina,
\newblock Reflection factors and a two-parameter family of boundary bound
  states in the sinh-Gordon model,
\newblock J. Phys. {\bf A33}, 8739--8754 (2000).

\bibitem{sinh7}
M.~Ablikim and E.~Corrigan,
\newblock On the perturbative expansion of boundary reflection factors of the
  supersymmetric sinh-Gordon model,
\newblock Int. J. Mod. Phys. {\bf A16}, 625--640 (2001).

\bibitem{Oota:1997un}
T.~Oota,
\newblock $q$-deformed Coxeter element in non-simply laced affine Toda field
  theories,
\newblock Nucl. Phys. {\bf B504}, 738--752 (1997).

\bibitem{frenkel}
E.~Frenkel and N.~Reshetikhin,
\newblock Deformations of W-algebras associated to simple Lie algebras,
  q-alg/9708006.

\end{thebibliography}

\end{document}